\documentclass[aps,prd,preprint,showpacs,preprintnumbers,nofootinbib]{revtex4}
\usepackage{epsfig}
\usepackage{latexsym,amsmath,amssymb,amstext}
\usepackage{float}
\usepackage{graphicx,xcolor}
\usepackage[paperwidth=210mm,paperheight=297mm,centering,hmargin=2cm,vmargin=2.5cm]{geometry}

\newcommand{\be}{\begin{equation}}
\newcommand{\ee}{\end{equation}}
\newcommand{\bea}{\begin{eqnarray}}
\newcommand{\eea}{\end{eqnarray}}

\newcommand{\tr}{{\rm tr\ }}

\newcommand\bef{\begin{figure}}
\newcommand\eef[1]{\label{fg:#1}\end{figure}}
\newcommand\beq{\begin{equation}}
\newcommand\eeq[1]{\label{#1}\end{equation}}
\newcommand\beqa{\begin{eqnarray}}
\newcommand\eeqa[1]{\label{#1}\end{eqnarray}}
\newcommand\bet{\begin{table}}
\newcommand\eet[1]{\label{tb:#1}\end{table}}

\newcommand\fgn[1]{Figure \ref{fg:#1}}
\newcommand\eqn[1]{Eq.\ (\ref{#1})}
\newcommand\scn[1]{Section \ref{sec:#1}}
\newcommand\apx[1]{Appendix \ref{sec:#1}}
\newcommand\tbn[1]{Table \ref{tb:#1}}

\newcommand\ie{{\sl i.e.\/}}

\begin{document}
\title{Scale-invariance of parity-invariant three-dimensional QED}
\author{Nikhil\ \surname{Karthik}}
\email{nkarthik@fiu.edu}
\affiliation{Department of Physics, Florida International University, Miami, FL 33199.}
\author{Rajamani\ \surname{Narayanan}}
\email{rajamani.narayanan@fiu.edu}
\affiliation{Department of Physics, Florida International University, Miami, FL 33199.}

\begin{abstract}
We present numerical evidences using overlap fermions for a
scale-invariant behavior of parity-invariant three-dimensional QED
with two flavors of massless two-component fermions.  Using finite-size
scaling of the low-lying eigenvalues of the massless anti-Hermitian
overlap Dirac operator, we rule out the presence of bilinear
condensate and estimate the mass anomalous dimension.  The eigenvectors
associated with these low-lying eigenvalues suggest critical behavior
in the sense of a metal-insulator transition.  We show that there
is no mass gap  in the scalar and vector correlators in the infinite
volume theory.  The vector correlator does not acquire an anomalous
dimension. The anomalous dimension associated with the long-distance
behavior of the scalar correlator is consistent with the mass
anomalous dimension.

\end{abstract}

\date{\today}
\maketitle

\section{Introduction}

Two-component massless fermions coupled to a three-dimensional
Euclidean abelian gauge field has been a topic of study in the past
three decades for several field-theoretic reasons.  The presence
of parity
anomaly~\cite{Deser:1981wh,Deser:1982vy,Redlich:1983dv,Niemi:1983rq}
induces a topological mass term for the gauge fields.  Soon after
that, it found an application in condensed matter physics as a
possible explanation of the quantum Hall effect~\cite{Semenoff:1984dq}.
Recently, duality between various theories in three dimensions that
includes fermions coupled to abelian gauge fields with or without
Chern-Simons matter are being discussed in the context of condensed
matter physics~\cite{Seiberg:2016gmd,Karch:2016sxi}.  Of particular
interest to us in this paper is the possible conformal nature of
parity-invariant theories with even number, $2N_f$, of massless
flavors.  This could have implications in the conductivity of
graphene type materials~\cite{Miransky:2015ava,CastroNeto:2009zz}.
Also, the $N_f=2$ theory seems to be of interest in the context of
high-$T_c$ cuprates~\cite{Franz:2002qy,Herbut:2002yq}.

A simple analysis of the associated gap equation~\cite{Pisarski:1984dj}
suggested that fermions could generate a mass as the number of
flavors tends to infinity. Subsequent analysis of the gap equation~\cite{
Appelquist:1985vf, Appelquist:1986fd, Appelquist:1986qw,
Appelquist:1988sr} reached an opposite conclusion that the infra-red
behavior of the large-$N_f$ theory is scale-invariant due to the
presence of a non-trivial fixed point. However, it also lead to the
possibility of a non-zero bilinear condensate if $N_f < 4$, and the
conclusion remained stable when $1/N_f$ correction was
included~\cite{Kotikov:2016wrb}.  A computation of renormalization
group flow including the presence of parity-invariant four-fermion
terms, lead to a critical number of flavors in the region of $N_f=4$
to $N_f=10$~\cite{Braun:2014wja}.  Comparing the free energies in
the IR and UV assuming non-interacting particles, one finds that
symmetry breaking is not expected when $N_f >
3/2$~\cite{Appelquist:1999hr,Appelquist:2004ib}.  Recently, there
has been a renewed interest in parity-invariant QED$_3$ due to the
presence of Wilson-Fisher fixed point in $4-\epsilon$ dimensions.
A similar comparison of the free energies, now assuming a conformal
phase and a broken phase, suggests that symmetry is not broken when
$N_f>4$~\cite{Giombi:2015haa}.  A computation~\cite{Giombi:2016fct}
of the coefficient of the two-point function of the stress energy
tensor in the $4-\epsilon$ expansion supports a conformal phase if
$N_f > 1 + \sqrt{2}$.  Computations~\cite{DiPietro:2015taa,Herbut:2016ide}
of the scaling dimensions of the naively irrelevant four-fermion
operators in the vicinity of the Wilson-Fisher fixed point suggest
that the four-fermion operators become relevant for $N_f < 2$, and
hence the possibility that the infra-red fixed point becomes unstable
for $N_f<2$.  A computation~\cite{Chester:2016ref} of the scaling
dimensions of the parity-even four-fermion operators in a $1/N_f$
expansion, taking into account the mixing with a larger basis of
operators, suggests that theories with $N_f > 1$ are conformal.

Earlier numerical work that studied the behavior of the fermion
bilinear as a function of the fermion mass using staggered
fermions~\cite{Hands:2002dv,Hands:2004bh} indicated that there is
evidence for a bilinear condensate for $N_f=1$. The evidence for a
condensate in $N_f=2$ was found to be weak. A numerical
study~\cite{Raviv:2014xna} of the beta function for $N_f=2$ theory
with Wilson fermions indicated that this is theory is not probably
conformal \footnote{We think that one can use the data presented
in Table-II of~\cite{Raviv:2014xna} and reach a conclusion that
$N_f=2$ theory has an IR fixed point.  The value of physical size
of the box, $\ell$ as defined in this paper, corresponding to the
values of $\beta$ and $L$ in~\cite{Raviv:2014xna} is $\ell=2L/\beta$.
A linear behavior can be seen in a plot of the inverse of the
dimensionless renormalized coupling, $1/g^2$, versus $1/\ell$ that
includes their data from all $L$.  A non-zero intercept at $1/\ell=0$
seen in their data suggests the existence of the IR
 fixed point at $g^2_c\approx 48$.}.  A recent
study~\cite{Karthik:2015sgq} of the spectrum of the low-lying
eigenvalues of the massless Wilson-Dirac operator did not show any
evidence for condensate for $N_f \ge 1$.

Of particular relevance is the U$(2N_f)$ global symmetry formally
present in the continuum fermion action
\be
S_f = \int d^3 x \sum_{i=1}^{2N_f} \bar\chi_i(x) \left[ \sum_{k=1}^3 \sigma_k \left\{ \partial_k +i A_k(x) \right\} \right] \chi_i(x),
\ee
where $\chi_i$ are the two-component fermion fields.  The U$(2N_f)$
symmetry is broken to U$(N_f)\times$U$(N_f)$ if one uses Wilson
fermions as a regulator on the lattice, and it is only recovered
in the continuum limit for massless fermions. Indeed, the numerical
computations in~\cite{Karthik:2015sgq} were done such that the
continuum limit was taken at a fixed physical volume and the infinite
volume limit was subsequently studied. As we will show in this
paper, the U$(2N_f)$ symmetry is present at the lattice level if
one uses overlap fermions. This is also the case if one regulates
using domain wall fermions~\cite{Hands:2015dyp,Hands:2015qha} and
take the limit of infinite number of fermions in the extra direction.

Since the overlap formalism in odd
dimensions~\cite{Narayanan:1997by,Kikukawa:1997qh} is not as well
known as in even dimensions, we start with an introduction to overlap
fermions in \scn{overlap}. We point out the U$(2N_f)$ symmetry
present at the level of the generating functional for massless
overlap fermions in a gauge field background. We perform numerical
simulations using massless overlap fermions for the case of $N_f=1$
and extract continuum results in a periodic box of size, $\ell^3$.
Our aim is to show that the $N_f=1$ theory is scale-invariant. We
explore three aspects of the theory to establish scale invariance:
\begin{enumerate}
\item 
If the low-lying eigenvalues $\lambda$ of the massless anti-Hermitian
overlap operator depend on the finite physical size $\ell$ as
\be
\lambda \sim \ell^{-1-\gamma_m}
\ee
with $\gamma_m<2$, then there is no bilinear condensate in the infinite 
volume theory and
the exponent $\gamma_m$ is the mass anomalous dimension. We
will show that $\gamma_m = 1.0 \pm 0.2$ in \scn{massdim}. This
is at the edge of the maximum allowed value for $\gamma_m$ in a
theory which is also conformally invariant~\cite{Mack:1975je}. 

\item In the sense of a metal-insulator 
transition~\cite{Osborn:1998nf,Altschuler:1986zh,Altschuler:1988al,Chalker:1996kr},
we will show that the eigenvectors associated with the low lying
eigenvalues lie in the critical regime. The inverse participation
ratio (IPR) of the eigenvectors $\Psi_\lambda$ is defined as
\beq
I_2= \frac{\int \left\{ \Psi^*_\lambda(x) \Psi_\lambda(x)\right\}^2 d^3 x}{\int \Psi^*_\lambda(x) \Psi_\lambda(x) d^3 x}.
\eeq{IPRdef}
In the critical regime, the IPR would exhibit a scaling with the physical
size $\ell$ as
\be
I_2 \sim \ell^{-3+\eta}.
\ee
This scaling is related to the behavior of the number variance $\Sigma_2(n)$,
the variance of the number of eigenvalues, $n$, below a given
value, $\lambda$. In the critical regime, $\Sigma_2(n)$ would
exhibit an asymptotic linear behavior with a slope $\eta/6$:
\be
\Sigma_2(n) = \frac{\eta}{6} n.
\ee
We will demonstrate in \scn{critical} that the low-lying eigensystem
of the $N_f=1$ theory satisfy such a critical behavior with $\eta=0.38(1)$.
\item We study the correlators of parity-even vector bilinear,
\be
V_k(x) = \bar\chi_1(x) \sigma_k \chi_1(x) - \bar\chi_2(x) \sigma_k \chi_2(x),
\ee
and the scalar bilinear 
\be
\Sigma(x) = \bar\chi_1(x) \chi_1(x)- \bar\chi_2(x) \chi_2(x),
\ee
in \scn{correlators}.  In both the cases, we will show there is no mass gap
in their spectrum, and that the long-distance behavior of the correlators
at zero spatial momentum exhibits a  power-law.
The power-law associated with the vector correlator does not acquire
any anomalous dimension consistent with the vector bilinear being a
conserved current.  The scalar correlator does not show a simple
power-law behavior as a function of Euclidean time, leading us to
estimate the expected power-law at even longer distances inaccessible
to our numerical simulation. The resulting value for the mass anomalous dimension
is $\gamma_m=0.8(1)$ which is consistent with the result from the low lying eigenvalues
and consistent with a vanishing correlator in the long distance limit.

\end{enumerate}

\section{Overlap formalism in three dimensions}\label{sec:overlap}

The overlap formalism for two-component fermions in three dimensions
were originally discussed in~\cite{Narayanan:1997by,Kikukawa:1997qh}
and more recently for parity-invariant four-component fermions by
starting from domain wall fermions~\cite{Hands:2015dyp,Hands:2015qha}.
In this paper, we start from the original overlap
formalism~\cite{Narayanan:1994gw} to obtain the result in three
dimensions. In this manner, we will explicitly show the
parity-invariant factorization into two two-component fermions.

\subsection{Gauge-invariant but parity-breaking overlap operator for a single flavor of 
two-component massless fermion}
The overlap formula~\cite{Narayanan:1994gw} for a two-component fermion determinant in three dimensions is 
\be
\det C_o = \langle 0 - | 0 + \rangle,
\ee
where
$|0\pm \rangle$ are the lowest states of the many body operators, 
\be
{\cal H}_\pm = -\begin{pmatrix}a^\dagger & b^\dagger\cr\end{pmatrix} H_\pm \begin{pmatrix} a \cr b \cr\end{pmatrix},
\ee
with $(a^\dagger,b^\dagger)$ and $(a,b)$ being two-component fermion creation and
annihilation operators that obey canonical anti-commutation relations.
The two single particle Hamiltonians are
\be
H_+  = \begin{pmatrix} B  & D  \cr -D  & -B \end{pmatrix};\qquad H_- = \gamma_5 = \begin{pmatrix} 1 & 0 \cr 0 & -1 \cr\end{pmatrix}.\label{sinpart}
\ee
The na\"ive massless Dirac operator in three dimensions is 
\be
D = \frac{1}{2} \sum_{k=1}^3 \sigma_k \left( T_k - T_k^\dagger\right);\qquad \left( T_k\phi\right)(x) = U_k(x) \phi(x+\hat k);\ \ \ \ 
T_k^\dagger T_k = 1.\label{naive}
\ee
with $\sigma_k$; $k=1,2,3$ being the two-component Pauli matrices.
The standard Wilson term is
\be
B =\frac{1}{2} \sum_{k=1}^3 \left( 2 - T_k - T_k^\dagger\right)-m_w;\ \ \ \
B=B^\dagger,\label{wilson}
\ee
with a Wilson mass parameter in the range $0 < m_w < 2$.

The na\"ive massless Dirac operator in three dimensions is
anti-Hermitian; $D^\dagger=-D$. Due to this special structure of
$H_+$ in \eqn{sinpart}, one can obtain an expression for $\langle
0 -| 0+ \rangle$ in odd dimensions in terms of an explicit operator.
Defining
\be
X = B+D,
\ee
we can set up the following eigenvalue problems:
\be
X^\dagger X \mathcal{R} = \mathcal{R} \Lambda^2;\qquad
X X^\dagger \mathcal{L} =\mathcal{L} \Lambda^2;\qquad
\Lambda_{ij} =\lambda_i \delta_{ij};\qquad \lambda_i > 0.\label{ebasis}
\ee
It follows that
\be
X =  \mathcal{L} \Lambda \mathcal{R}^\dagger;\qquad
V \equiv \mathcal{L}\mathcal{R}^\dagger = X \frac{1}{\sqrt{X^\dagger X}} = \frac{1}{\sqrt{XX^\dagger}} X;\qquad VV^\dagger=1.\label{voper}
\ee
The basis of positive and negative eigenstates of $H_+$ are
\be
\frac{1}{2}\begin{pmatrix} (1+V)\mathcal{R}\cr (1-V) \mathcal{R} \cr \end{pmatrix}; \qquad
\frac{1}{2}\begin{pmatrix} (1-V) \mathcal{R} \cr (1+V) \mathcal{R}\cr \end{pmatrix};
\label{evhpmr}
\ee
respectively.
On the one hand, the basis of positive and negative eigenstates of $H_-$ can be chosen to be
\be
\begin{pmatrix} \mathcal{R}\cr 0\cr \end{pmatrix}; \qquad
\begin{pmatrix} 0\cr \mathcal{R}\cr \end{pmatrix};
\label{evhmmr}
\ee
respectively and the fermion determinant for a single two-component fermion becomes
\be
\det C_o = \det \frac {1+V}{2}.
\ee
On the other hand, the basis of positive and negative eigenstates of $H_-$ can be chosen to be
\be
\begin{pmatrix} \mathcal{L}\cr 0\cr \end{pmatrix}; \qquad
\begin{pmatrix} 0\cr \mathcal{L}\cr \end{pmatrix};
\ee
respectively and the fermion determinant for a single two-component fermion becomes
\be
\det C_o = \det \frac {1+V^\dagger}{2}.
\ee
In both cases, the fermion determinant is gauge
invariant~\cite{Kikukawa:1997qh}. The phase of the fermion determinant
arises from the phase of the Wilson-Dirac operator $X$ analyzed in
detail in~\cite{Karthik:2015sza} and carries the parity
anomaly~\cite{Narayanan:1997by,Kikukawa:1997qh}.

\subsection{U$(2N_f)$ symmetric parity and gauge-invariant overlap formalism}
In order to realize a parity-invariant theory, we consider theories
with even number of fermion flavors, $2N_f$.  As shown in \apx{z2mass},
the result for the generating function for a $N_f=1$ theory, including
parity invariant fermion masses and using flavor diagonal sources $\eta_+$ and $\eta_-$,
is
\be
Z_2(\eta_+,\eta_-,\bar\eta_+,\bar\eta_-;m) = \left\{\det C_o(m) 
\exp \left[ \bar\eta_+ G_o(m) \eta_+\right]\right\}
\left\{\det C_o^\dagger(m) \exp\left[ - \bar\eta_- G_o^\dagger(m) \eta_- 
\right]\right\},\label{z2gen}
\ee
where
\be
C_o(m) = \frac{1+V}{2} + m \frac{1-V}{2};\qquad G_o(m) = \frac{1}{1-m} \left[ C_o^{-1}(m) -1 \right] = \frac{A}{1+mA},
\ee
and
\be
A = \frac{1-V}{1+V}.
\ee
If we set $m=0$, we
observe that $G_o(0) = -G_o^\dagger(0) = A$. Therefore, one has the
full U$(2N_f)$ symmetry in the fermionic sources. The fermion
determinant in the absence of sources for the massless $N_f=1$ theory is
\be
\det \frac{1+V}{2} \det \frac{1+V^\dagger}{2} =
\det V^\dagger \det \left[ \frac {1+V}{2}\right]^2.
\ee
This also shows a U$(2N_f)$ symmetry, but we have an additional factor
of $\det V^\dagger$ for each $N_f$ in the measure for the gauge fields,
which is required to keep the theory parity invariant.

\subsection{Eigenvalues of the overlap operator}
The bilinear scalar condensate in a fixed gauge field background is given
by
\be
\Sigma(m) = \frac{1}{2\ell^3}\int \langle \Sigma(x) \rangle d^3 x
= \frac{1}{2\ell^3} \tr\left[G_o(m) + G_o^\dagger(m) \right]
= \int_{-\infty}^\infty d \Lambda \frac{ \rho(\Lambda) }{i\Lambda+m},\quad \int_{-\infty}^\infty \rho(\Lambda) =1;
\ee
where $i\Lambda$ are the eigenvalues of $A^{-1}$ and $\rho(\Lambda)$
is the density of eigenvalues obeying $\rho(\Lambda)=\rho(-\Lambda)$.
Therefore, we are led to an analysis of the low-lying eigenvalues
of $A^{-1}$ in order to extract the mass anomalous dimension.

Given an eigenvalue $e^{i\Phi_i}$ of $V$, the corresponding eigenvalue of 
$A^{-1}$ is 
\be
i \Lambda_i = i \cot \frac{\Phi_i}{2}.
\ee
Therefore, the low-lying eigenvalues come from values of $\Phi_i$ close to
$\pi$. In order to obtain these numerically using the Ritz algorithm~\cite{Kalkreuter:1995mm},
we compute the low-lying eigenvalues of the positive definite operator,
 \be
 C_0(0) C_0^\dagger(0) = \frac{2+V+V^\dagger}{4}.\label{ccdag}
 \ee
The corresponding eigenvalue of this operator is $\cos^2\frac{\Phi_i}{2}$.
Due to parity invariance, we only need
\be
|\Lambda_i| = \frac{ |\cos \frac{\Phi_i}{2}|}{\sqrt{ 1- \cos^2\frac{\Phi_i}{2}}}.
\ee

\section{Set up of the numerical calculation}

The numerical details essentially parallel the one used in our
previous work~\cite{Karthik:2015sgq} with Wilson fermions. The only
new ingredient is the presence of the operator $V$ defined in
\eqn{voper}. We used a $21^{\rm st}$ order Zolotarev
approximation~\cite{vandenEshof:2002ms,Chiu:2002eh} to realize
$\frac{1}{\sqrt{X^\dagger X}}$ and this was sufficient for all our
simulation parameters.  We worked with massless fermions and the
pseudofermion operator was written as
 \be
 S_f(\phi) = \left( C_o^{-1}(0)\phi\right) ^\dagger C_o^{-1}(0) \phi
 = \phi^\dagger \frac{4}{2+V+V^\dagger} \phi.\label{pseudo}
 \ee
We would like to draw attention to the advantage of using two-flavors
of two component fermions instead of using an equivalent single
four-component fermion formalism; the fermion determinant is a
determinant of a positive-definite operator enabling the Monte Carlo
simulation for any value of $2N_f$.  We worked on a 3d torus of fixed
physical extent $\ell$ and regulated using a $L^3$ lattice.  Since
it is a bit different from the standard procedure, we note
that $L$ is used to tune the lattice spacing at a fixed physical
size, $\ell$.  We used $L=12,14,16,20$ and 24 to extract the continuum
limit of observables.  It is worth noting that unlike in four
dimensions, there could be $\mathcal{O}(1/L)$ corrections due to
the presence of parity-even, dimension-four, four-fermion operators
which preserve the U$(2N_f)$ flavor symmetry.  We used several
different values of $\ell$ to understand the behavior of
the theory as a function of $\ell$, and then properly obtain the
behavior as $\ell\to\infty$.  We list our simulation points in the
\apx{details}.

In order to take the continuum limit, one has to take into account
a factor arising from the Wilson mass parameter to realize the
correct dispersion relation for free fermions~\cite{Edwards:1998wx}.
At a finite lattice spacing, this factor can be improved and we
define the improved eigenvalues by
\beq
\lambda_i= Z_m\Lambda_i\quad;\quad Z_m=2(m_w-m_t),
\eeq{improved}
where $m_w$ is the mass used in the Wilson-Dirac kernel, and $m_t$
is the Wilson mass at which the smallest eigenvalue is minimum.  We
use $m_w=1$ at all simulation points. The values of $m_t$ are listed
in the \apx{details}.

We compute the correlators of scalar and vector bilinears 
at zero spatial momentum defined in the continuum by
\be
G_\Sigma(t)= \int dx dy \Big\langle \Sigma(x,y,t) \Sigma(0,0,0)\Big\rangle \quad\text{and}\quad G_V(t)= \int dx dy \Big\langle V_i(x,y,t) V_i(0,0,0)\Big\rangle,
\ee
respectively. On the lattice, after Wick contractions, the correlators using 
massless fermions become
\bea
G_\Sigma(T) &=&\frac{L^2}{\ell^2 Z^2_m} \sum_{X,Y}
\tr \big[ A(0,0,0;X,Y,T) A(X,Y,T;0,0,0) \big]\qquad\text{and}
\cr
G_V(T) &=&\frac{L^2}{\ell^2} \sum_{i=1}^2\sum_{X,Y}
\tr \big[ \sigma_i A(0,0,0;X,Y,T)\sigma_i  A(X,Y,T;0,0,0)\big],
\label{ops}
\eea
where ``$\tr$" denotes the trace over the spin index. One of the fermion
bilinear is placed at (0,0,0) and the other at (integer) lattice
coordinates $(X,Y,T)$ \ie, $T=t L/\ell$ and so on.  The factor $Z_m$
in the scalar propagator takes care of the renormalized scalar
operator $Z_m^{-1} \Sigma$.  Note that since $\Sigma$ and $V$ are
bosonic, their correlators are periodic functions of $T$ with period
$L$.  Therefore, we only show the correlators from $T=0$ to $T=L/2$
in all the plots in this paper.

\section{Mass anomalous dimension using $\ell$-scaling of low-lying eigenvalues}\label{sec:massdim}

\bef
\centering
\includegraphics[scale=0.55]{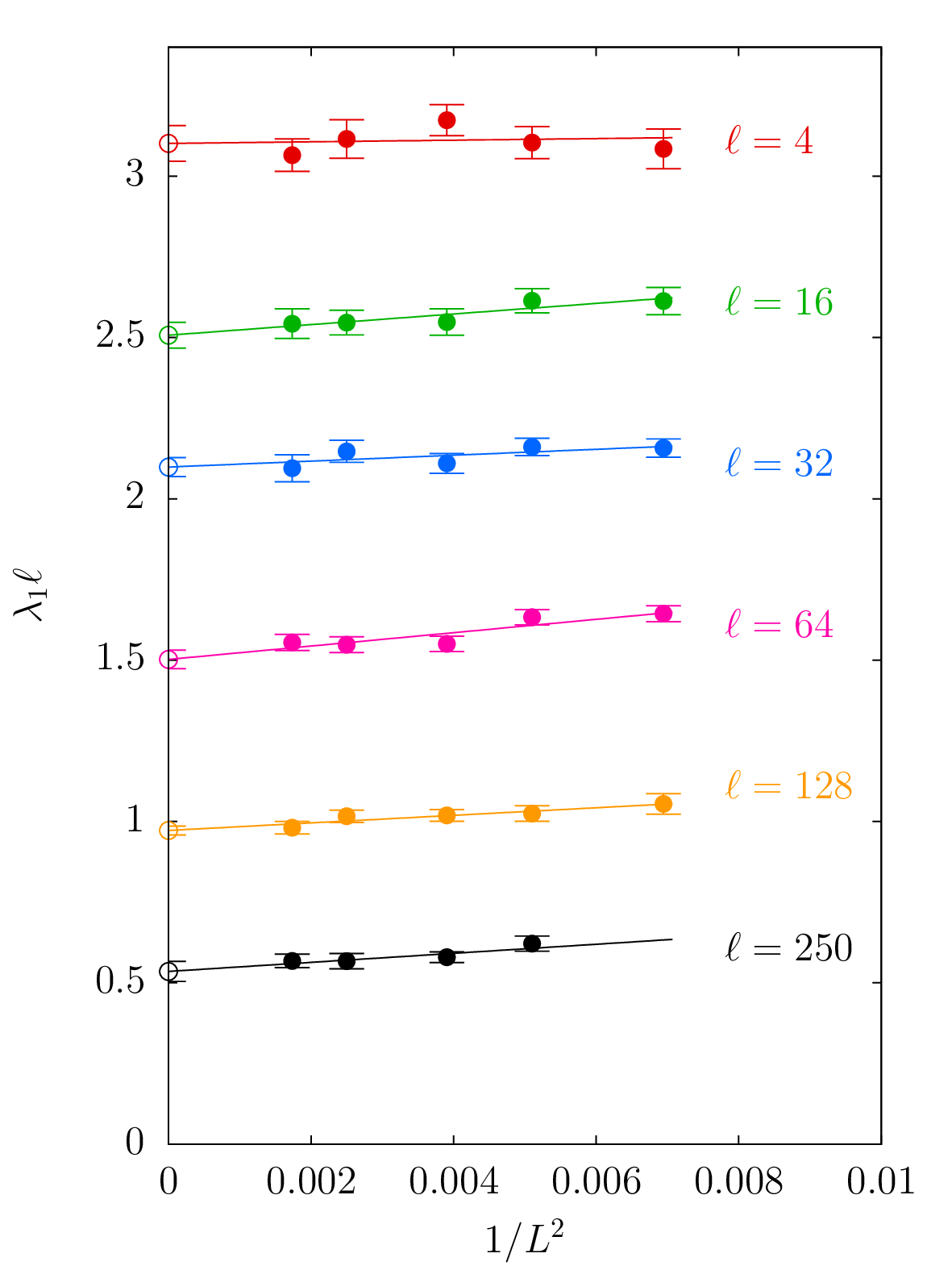}
\includegraphics[scale=0.55]{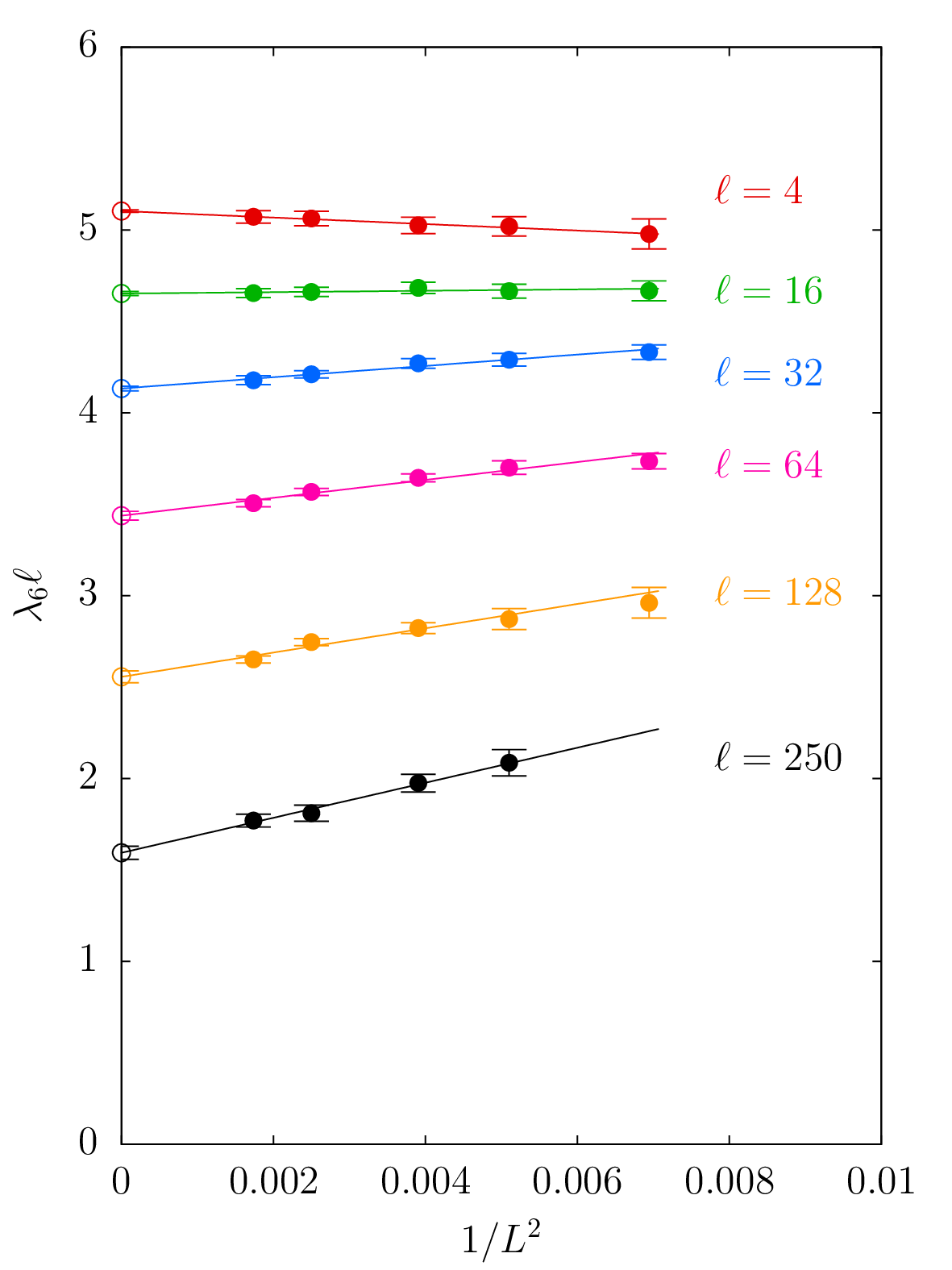}
\caption{The dimensionless improved eigenvalues $\lambda_i \ell$
at various $\ell$ is shown as a function of $1/L^2$ in order to
show the remaining dominant $1/L^2$ lattice artifact after the
improvement using $Z_m$.  The left and the right panels are for the
smallest $\lambda_1$ and a larger $\lambda_6$ respectively.  The
data points are from the simulations using
$L=12,14,16,20$ and 24 lattices.  The continuum limits are taken
using $1/L^2$ extrapolations, which are shown as the straight lines
in the plots.}
\eef{eigencont}

Our previous analysis~\cite{Karthik:2015sgq} using massless Wilson
fermions provided no evidence for a nonzero bilinear condensate and
we found $\lambda_i \sim \ell^{-2}$ for the $N_f=1$ theory.  If
we assume
\beq
\lambda_i \sim \frac{1}{\ell^{1+\gamma_m}},
\eeq{degrand}
and find $\gamma_m<2$,
it follows that the $\gamma_m$ is the mass anomalous dimension since
$\lambda_i$ has the dimensions of mass.  
In this section, we show that our
current simulations with overlap fermions produces results that are
quite consistent with our previous studies using massless Wilson
fermions for the case of $N_f=1$.
\bef
\centering
\includegraphics[scale=0.8]{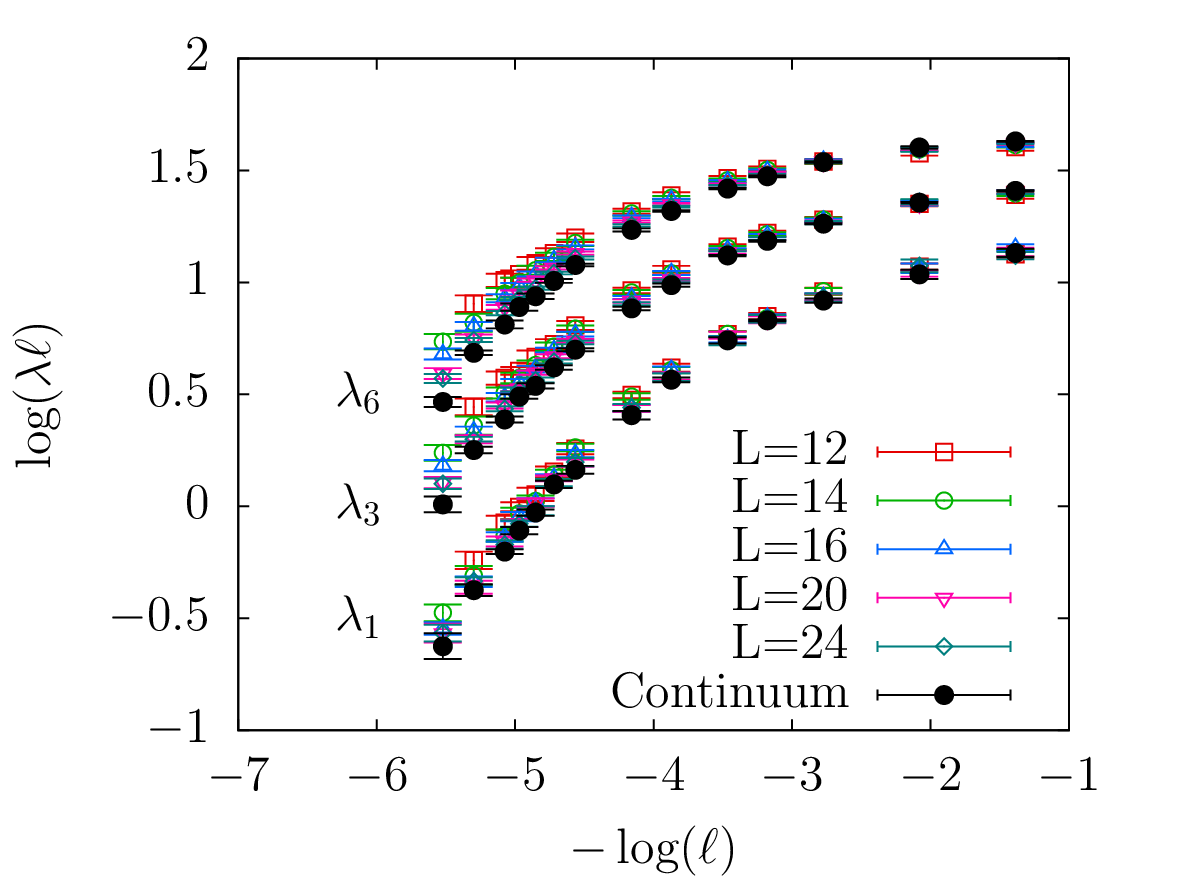}
\includegraphics[scale=0.8]{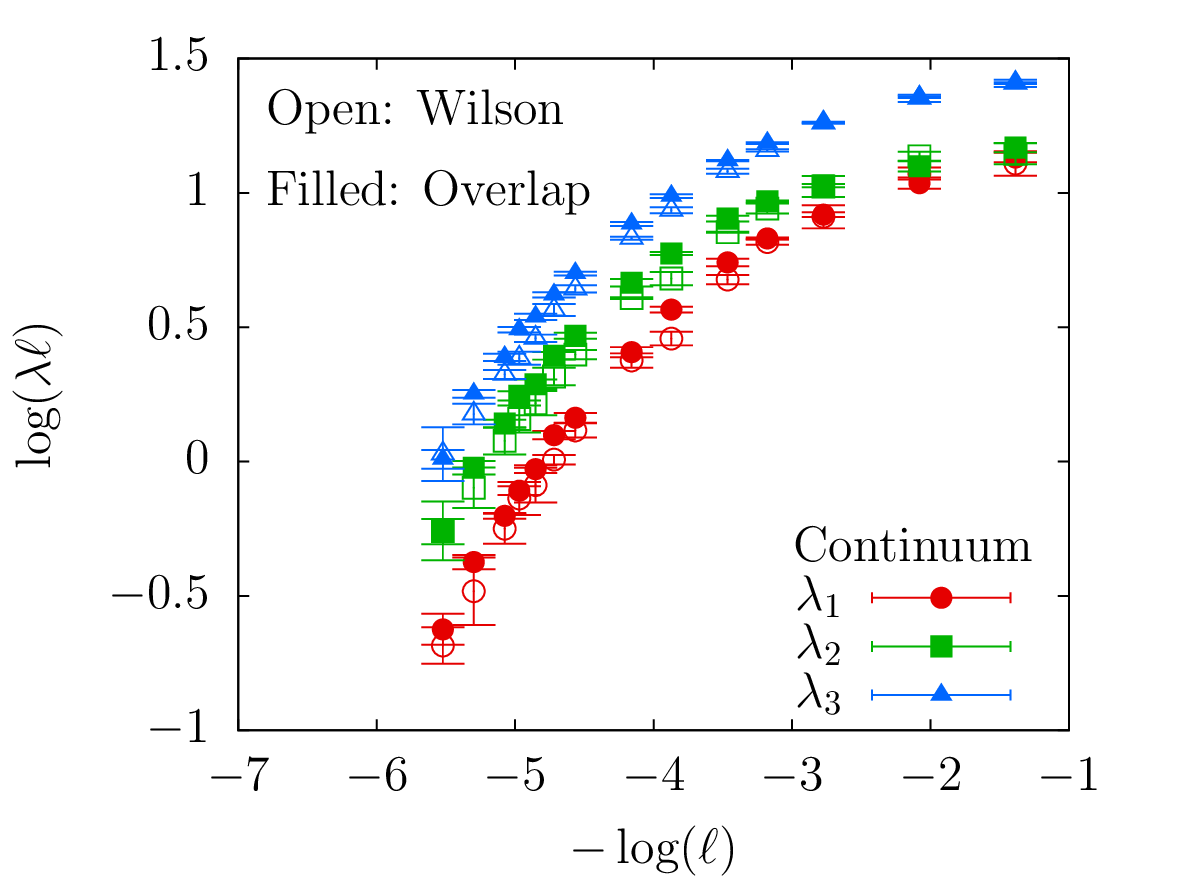}
\caption{The top panel shows representative eigenvalues in the
continuum along with the ones in finite lattice spacing, as a
function of $\ell$. On the bottom panel, the continuum limits using
overlap (filled symbols) and Wilson-Dirac fermions (open symbols)
are compared.}
\eef{eigvsell}

\bef
\centering
\includegraphics[scale=0.8]{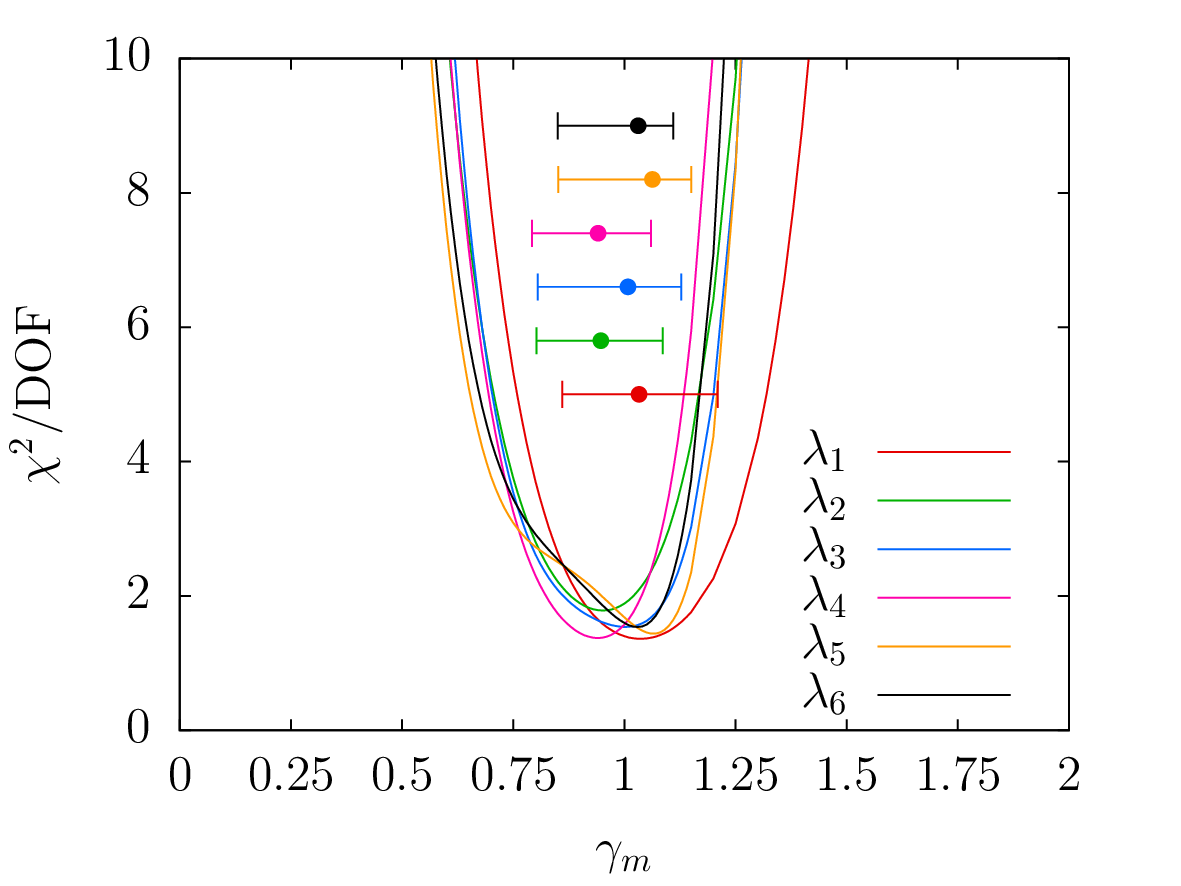}
\caption{
The $\chi^2/$DOF for the fits using the ansatz in \eqn{pade} to the
eigenvalue data is shown as a function of the exponent $\gamma_m$.
These are shown by the colored solid curves for the smallest six
eigenvalues. The corresponding 68\% confidence interval for the
exponent are shown by the error bars.
}
\eef{efitp}

We show the approach to the continuum limit for the improved first
and sixth positive eigenvalues, $\lambda_1$ and $\lambda_6$, at different fixed $\ell$
in \fgn{eigencont}. We find that the leading $\mathcal{O}(1/L)$
lattice corrections are removed by the factor $Z_m$. Using $1/L^2$
extrapolations, we obtain the continuum limit of the eigenvalues.
On the  top panel of \fgn{eigvsell}, we show the continuum limit
so obtained as a function of $\ell$, along with the eigenvalues at
finite $L$.
On the bottom
panel of \fgn{eigvsell}, we compare the continuum limits obtained
using overlap fermions with our earlier result using massless Wilson
Dirac operator~\cite{Karthik:2015sgq}.  A good agreement between
the two lattice regularizations is seen.

We do not find a simple power-law scaling in the region of $\ell$
where we simulated.  We only know the asymptotic dependence of
$\lambda$ on $\ell$; we expect the eigenvalues to behave proportional
to $\frac{1}{\ell}$ for small $\ell$ since the theory is asymptotically
free, and as $\frac{1}{\ell^{1+\gamma_m}}$ for large $\ell$. In
order to fit the data over the entire range of $\ell$, we found it
convenient to parametrize the dependence on $\ell$ in terms of
$\tau=\tanh(1/\ell)$.  Since we do not know the functional dependence
of $\lambda$ on $\ell$ at any intermediary $\ell$, we approximate
this functional dependence through a rational $[1/1]$ Pad\'e
approximant:
\beq
\lambda\ell = a_1 \tau^{-\gamma_m} \frac{1+a_2\tau}{1+a_3\tau},
\eeq{pade}
where the $a$'s are fit parameters.  The $\chi^2/{\rm DOF}$ as a
function of $\gamma_m$ is shown in \fgn{efitp}. All six low lying
eigenvalues predict a value of $\gamma_m=1.0\pm 0.2$ with $68\%$
confidence.  The fit to the data using $\gamma_m=1$ is shown in
\fgn{eigfit}.

\bef
\centering
\includegraphics[scale=0.75]{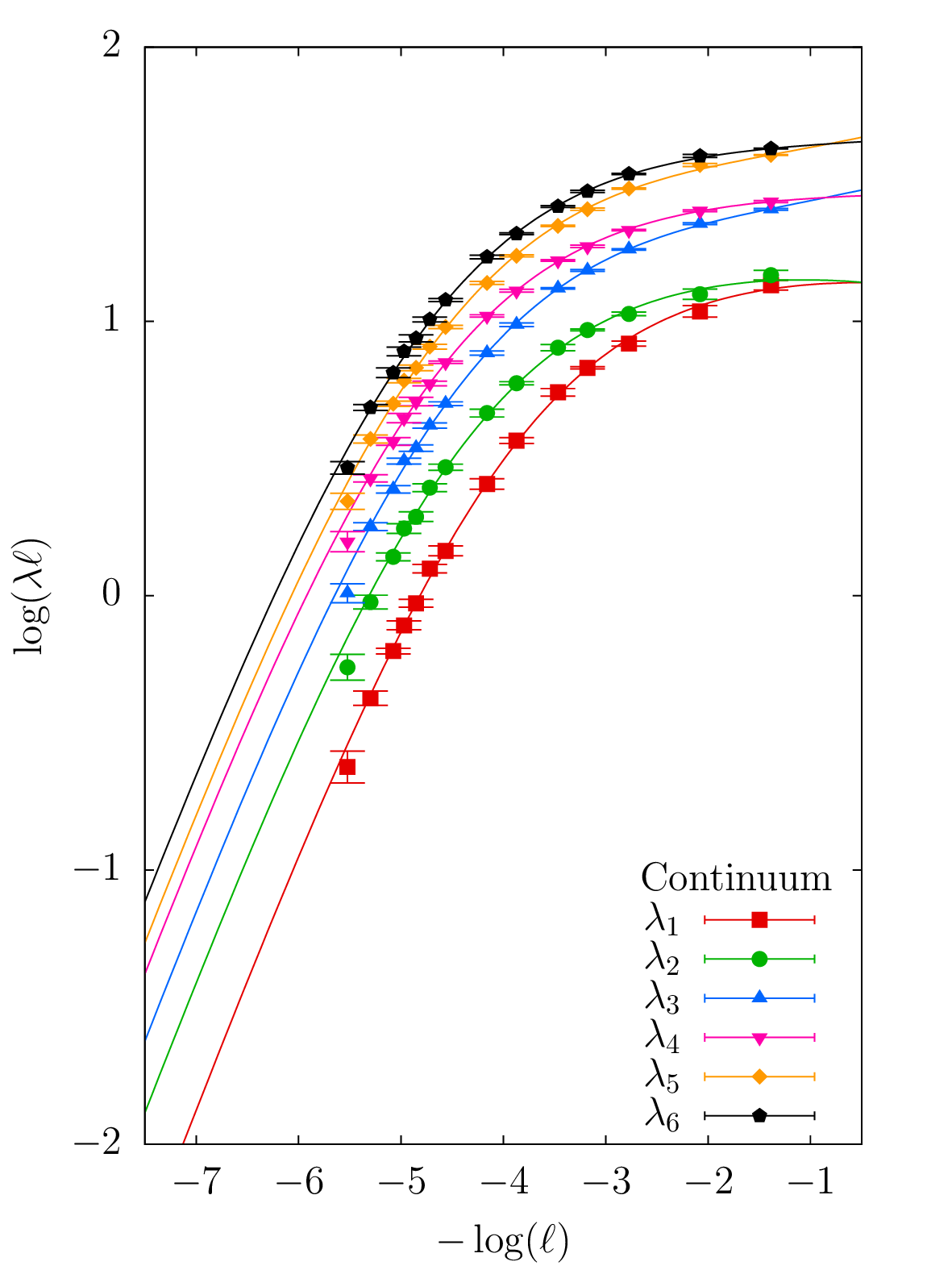}
\caption{ $\log(\lambda\ell)$ is shown as a function of $-\log(\ell)$
for the smallest six eigenvalues after taking their continuum limits.
A power-law $\lambda\sim\ell^{-\gamma_m-1}$ would be a straight
line with a slope $\gamma_m$ in this plot. No distinct power law
is seen in the volumes that we simulated. The $\ell$-dependence is
well described a $[1/1]$ Pad\'e approximant in \eqn{pade} using
which we estimate the eventual power-law behavior that would set
in at even larger $\ell$ than we used. The best fits using  $\gamma_m=1$
are shown by the solid curves.
}
\eef{eigfit}

\section{Inverse Participation Ratio and number variance}\label{sec:critical}
\bef
\centering
\includegraphics[scale=0.8]{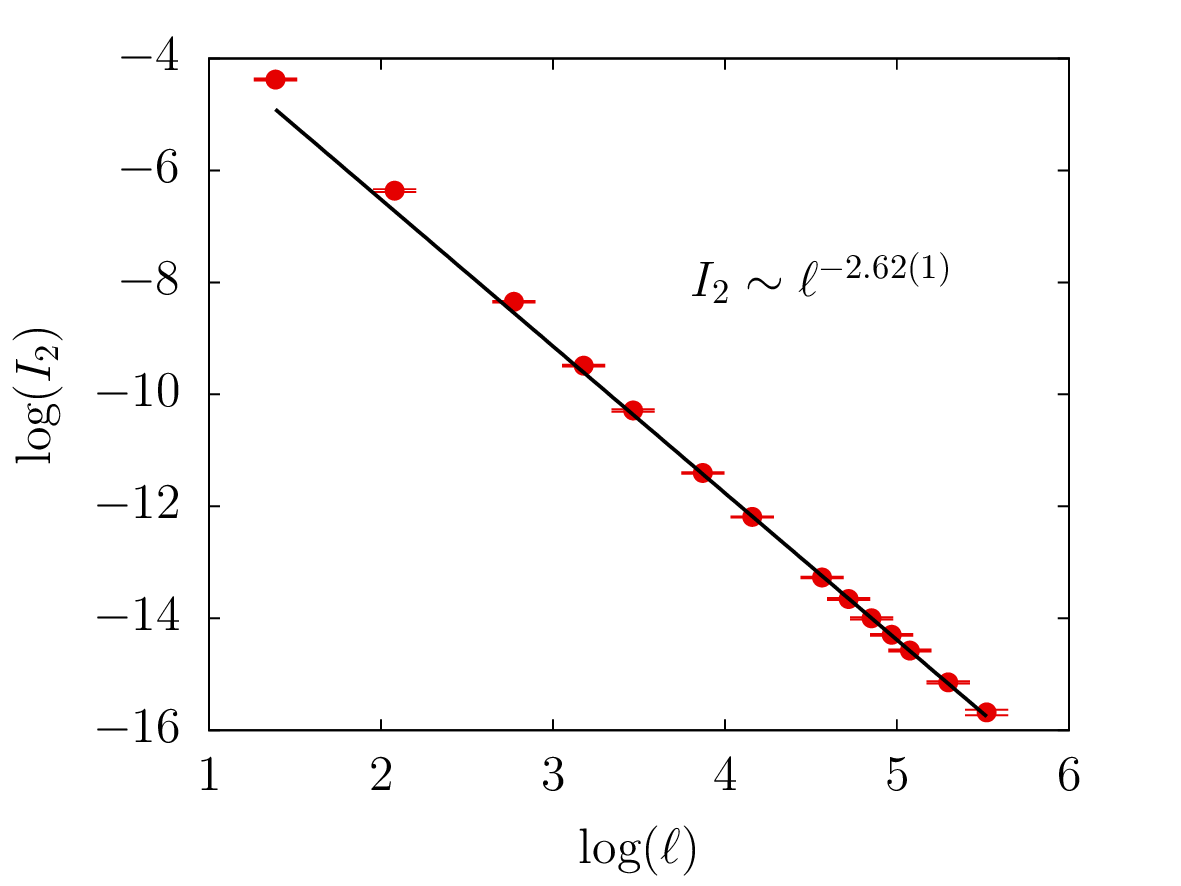}
\includegraphics[scale=0.8]{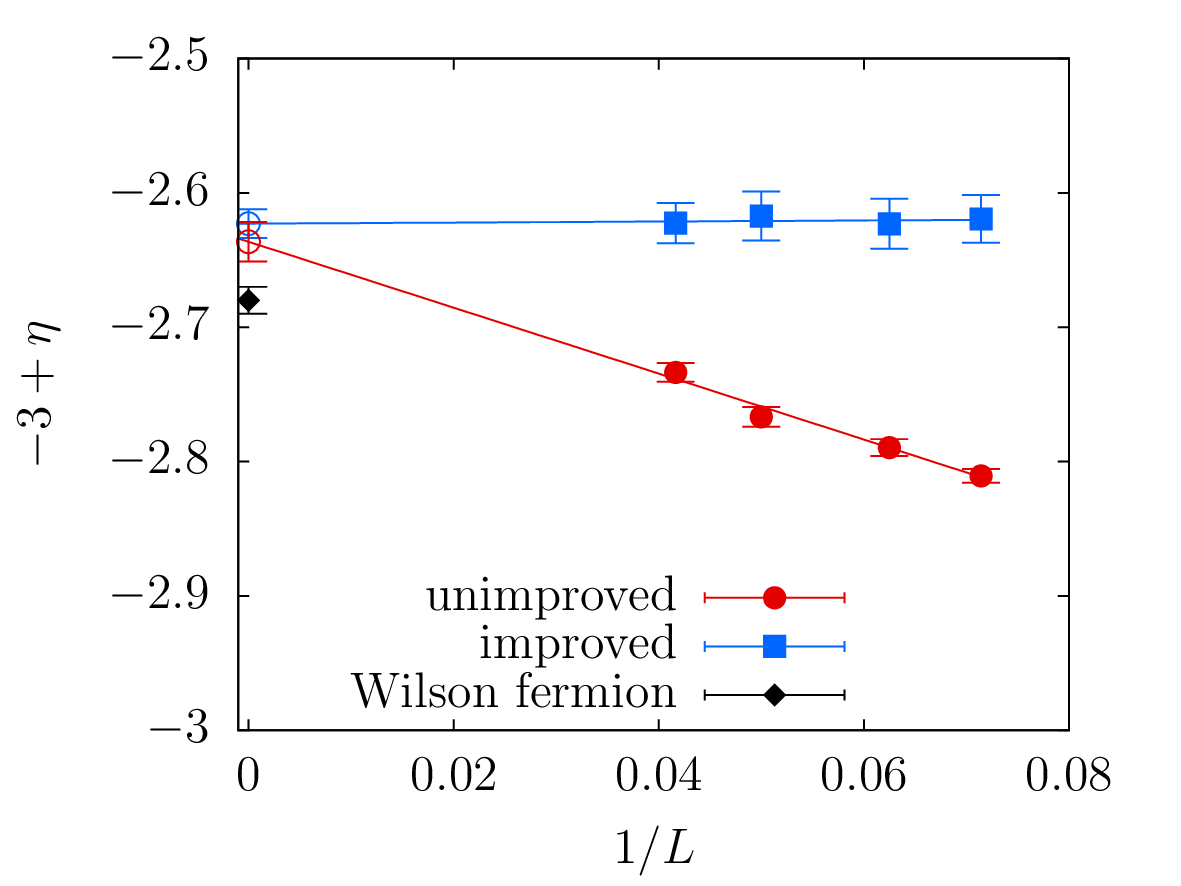}
\caption{
(top panel) The scaling $I_2\sim\ell^{-3+\eta}$ is shown using the
$L=24$ data. (Bottom panel) The continuum limit of the exponent
$-3+\eta$ using both $I_2$ (unimproved) and $I_2/Z_m$ (improved)
are shown. We estimate $-3+\eta=-2.62(1)$ in the continuum limit.
This is close to the value $-2.68(1)$  estimated using Wilson
fermions.
}
\eef{ipr}

\bef
\centering
\includegraphics[scale=0.7]{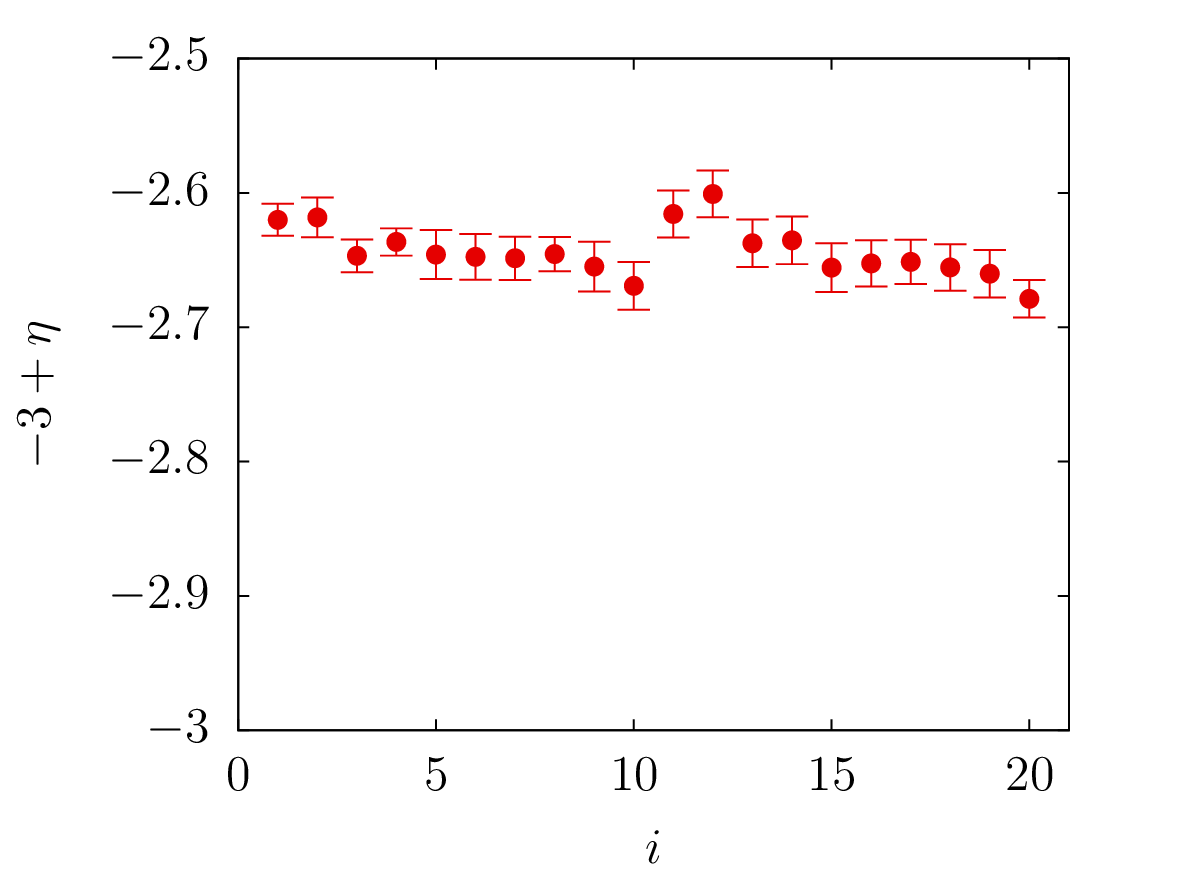}
\caption{
The exponent $-3+\eta$ for the finite-size scaling of the IPR for
various eigenvectors corresponding to the twenty low-lying eigenvalues
$\lambda_i$.
}
\eef{iprall}

\bef
\centering
\includegraphics[scale=0.70]{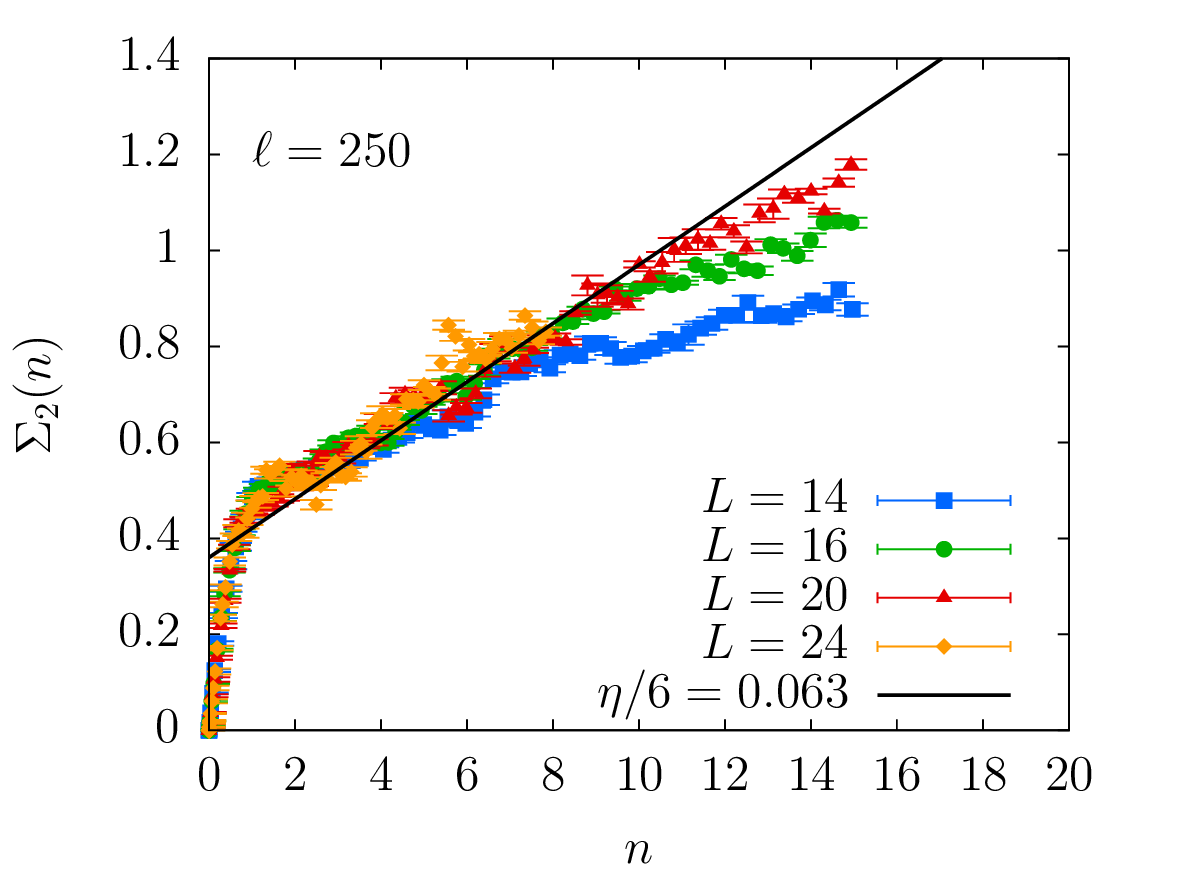}
\includegraphics[scale=0.70]{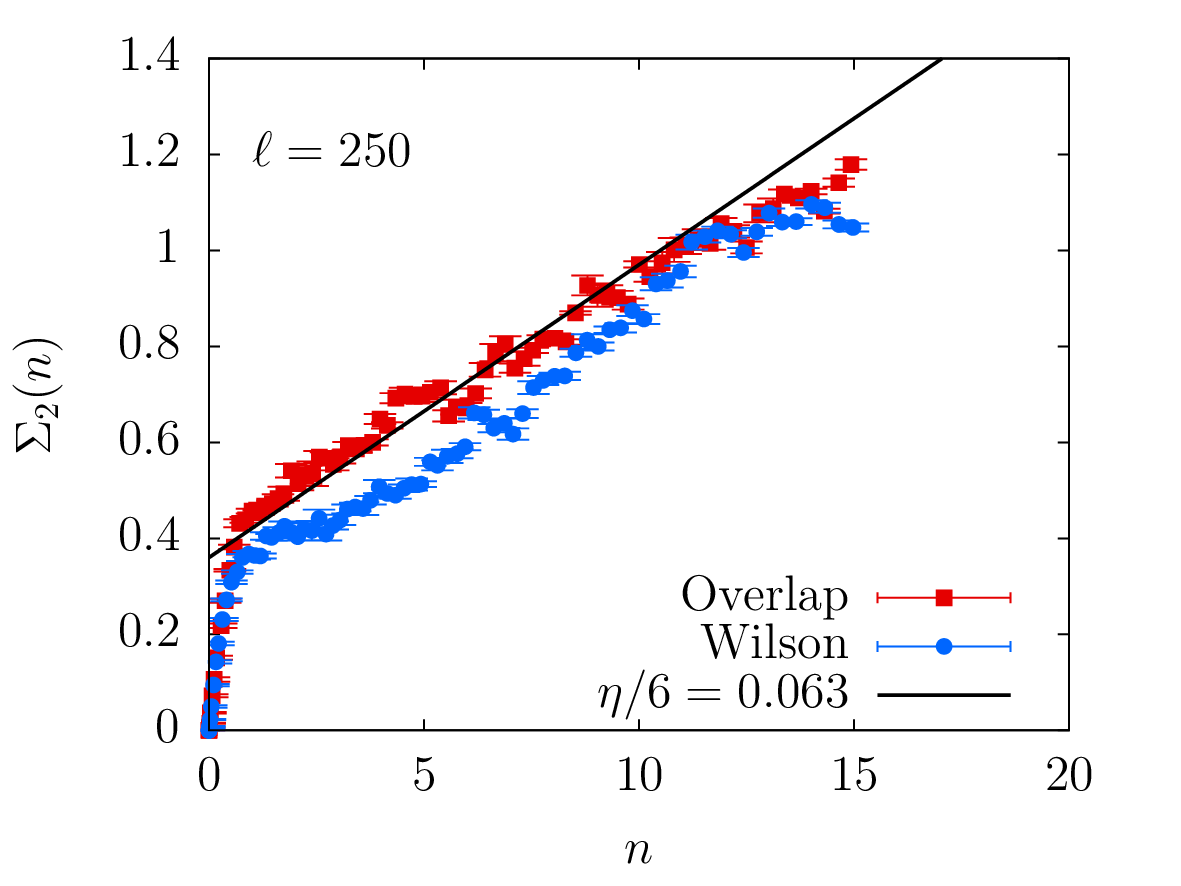}
\caption{
Lattice spacing effect in number variance. (Left)
Number variance $\Sigma_2(n)$ as a function of $n$ is shown for
$\ell=250$ at various lattice sizes $L$. As lattice spacing is
reduced, the number variance at larger $n$ increases and approaches the solid
black line which has a slope of $\frac{\eta}{6}=0.063$.
(Right) $\Sigma_2(n)$ for the overlap ($L=20$) and the Wilson-Dirac
fermions ($L=28$) at $\ell=250$ are compared. 
}

\eef{numvar}
\bef
\centering
\includegraphics[scale=0.8]{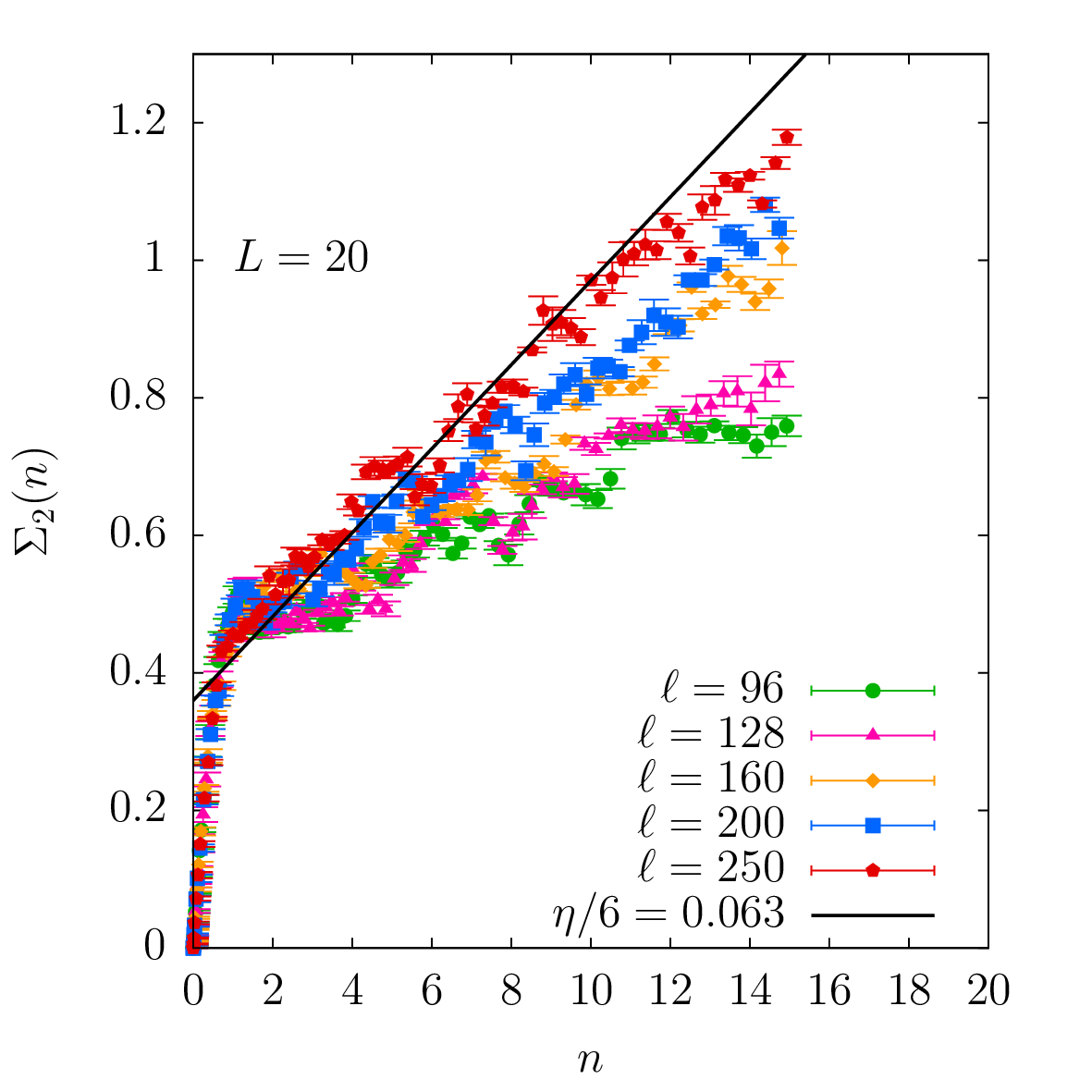}
\caption{
The number variance is shown as a function of $n$ for various $\ell$
on $L=20$ lattice. The black line has a slope of
$\frac{\eta}{6}=0.063$. The slope of the number variance increases with 
$\ell$ and approaches $\eta/6$ which is a trend opposite of that expected 
when a bilinear condensate is present.
}
\eef{numvarell}

In the absence of a bilinear condensate, there is no ergodic regime
in the eigenvalue spectrum of the massless overlap Dirac operator
similar to the observations made about the massless Wilson Dirac
operator in~\cite{Karthik:2015sgq}.  The ergodic regime is characterized
by eigenvectors that are completely delocalized and characterized
by an IPR defined in \eqn{IPRdef} which scales 
as $I_2 \sim \ell^{-3}$. A complete localization of the
eigenvectors will correspond to a value equal to $I_2 =1$. Instead,
we observe a power law behavior with $\ell$ that is consistent with
critical behavior~\cite{Osborn:1998nf,Altschuler:1986zh,Altschuler:1988al,Chalker:1996kr},
\be
I_2 \sim \ell^{-3+\eta},
\ee
with $\eta=0.38(1)$ in the continuum limit as shown in \fgn{ipr}.
The top panel of \fgn{ipr} shows the finite size scaling of $I_2$
for the eigenvector corresponding to the smallest eigenvalue, as
determined on $L=24$ lattice.  We find IPR to be one of the few
observables which show a simple power-law behavior over a range of
$\ell$ we simulated. In the bottom panel, we show the exponent
$-3+\eta$ as a function of $1/L$. The red solid circles are the
ones without any improvement, and it shows a leading $\mathcal{O}(1/L)$
lattice correction. The continuum extrapolated value corresponds
to the one we quoted: $\eta=0.38(1)$.  We empirically find the
$\mathcal{O}(1/L)$ to be removed by using an improved definition,
$I_2/Z_m$. These are shown by the blue solid square points, which
extrapolates to the same value of $\eta$. The black solid diamond
data point corresponds to the value as determined using Wilson-Dirac
fermions~\cite{Karthik:2015sgq}. In \fgn{iprall}, we show the
exponent $-3+\eta$ for the $i$-th eigenvector, for different
$i=1,2,\ldots, 20$.  We find the finite-size scaling of IPR to be
robust across eigenvectors. The small disagreement in $-3+\eta$
between the overlap and Wilson fermion, is not significant compared
to the scatter seen in \fgn{iprall}.

If the low-lying eigenvalues are in the critical regime, then we
expect the number variance, $\Sigma_2(n)$, to behave linearly with
$n$, with a slope given by $\frac{\eta}{6} = 0.063(3)$.  On the
left panel of \fgn{numvar}, we show the number variance as a function
of $n$ at $\ell=250$ at different $L$. We do see some finite lattice
spacing effects at larger $n$. The slope of $\Sigma_2$ from the
finer $L=24$ lattice matches the critical behavior for a wide range
of $n$.  On the right panel of \fgn{numvar}, we compare the result
from overlap fermion  with the one from the Wilson-Dirac
fermion~\cite{Karthik:2015sgq}.  The linear behavior is seen for a
wider range with overlap fermions.  Perhaps, this is because overlap
fermions are exactly massless thereby capturing the fluctuations
of the low-lying eigenvalues better than the Wilson-Dirac fermions.
Finally, we show the $\ell$-dependence of the number variance in
\fgn{numvarell} where we see that the slope of the linear rise
increases with $\ell$ and approaches the slope $\eta/6$. As we noted
in~\cite{Karthik:2015sgq}, this trend is opposite to the
one expected when a condensate is present.

\section{Scalar and vector correlators}\label{sec:correlators}

\bef
\centering
\includegraphics[scale=0.8]{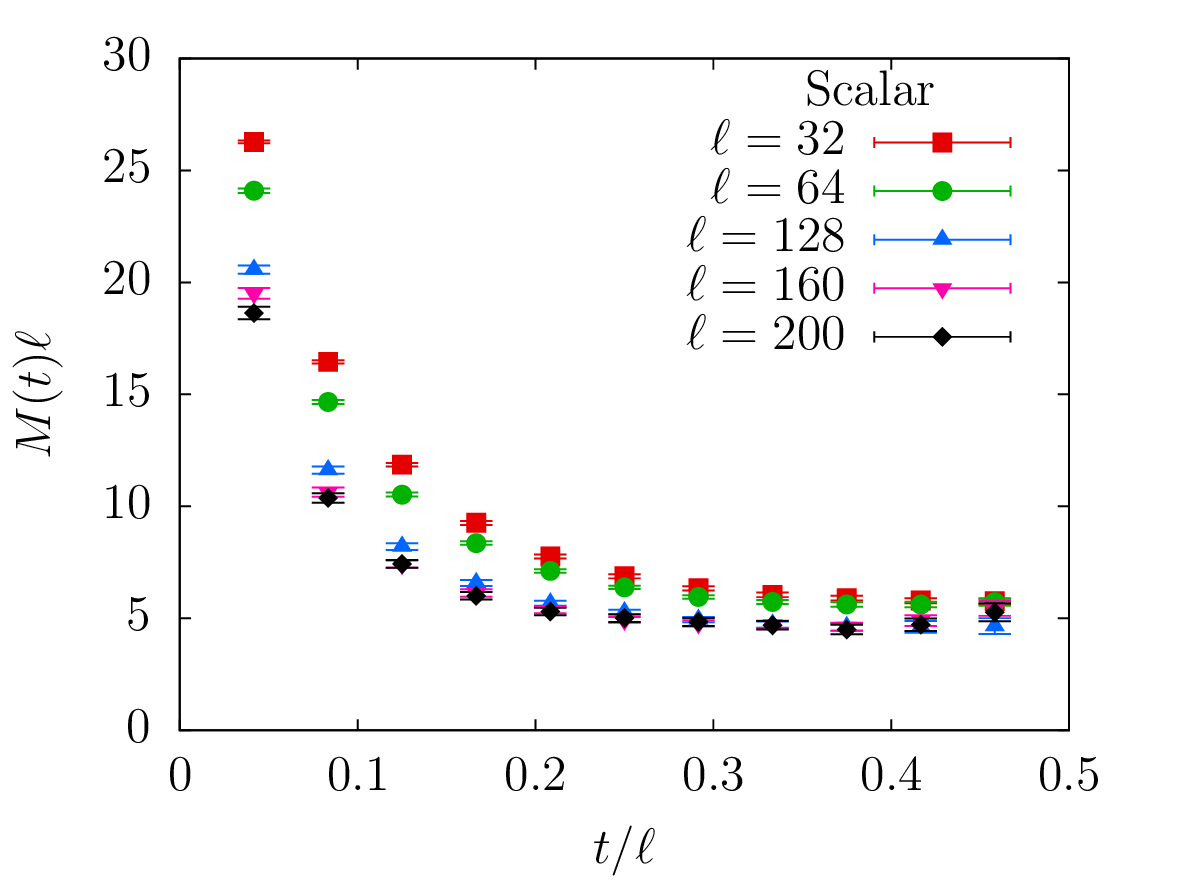}
\includegraphics[scale=0.8]{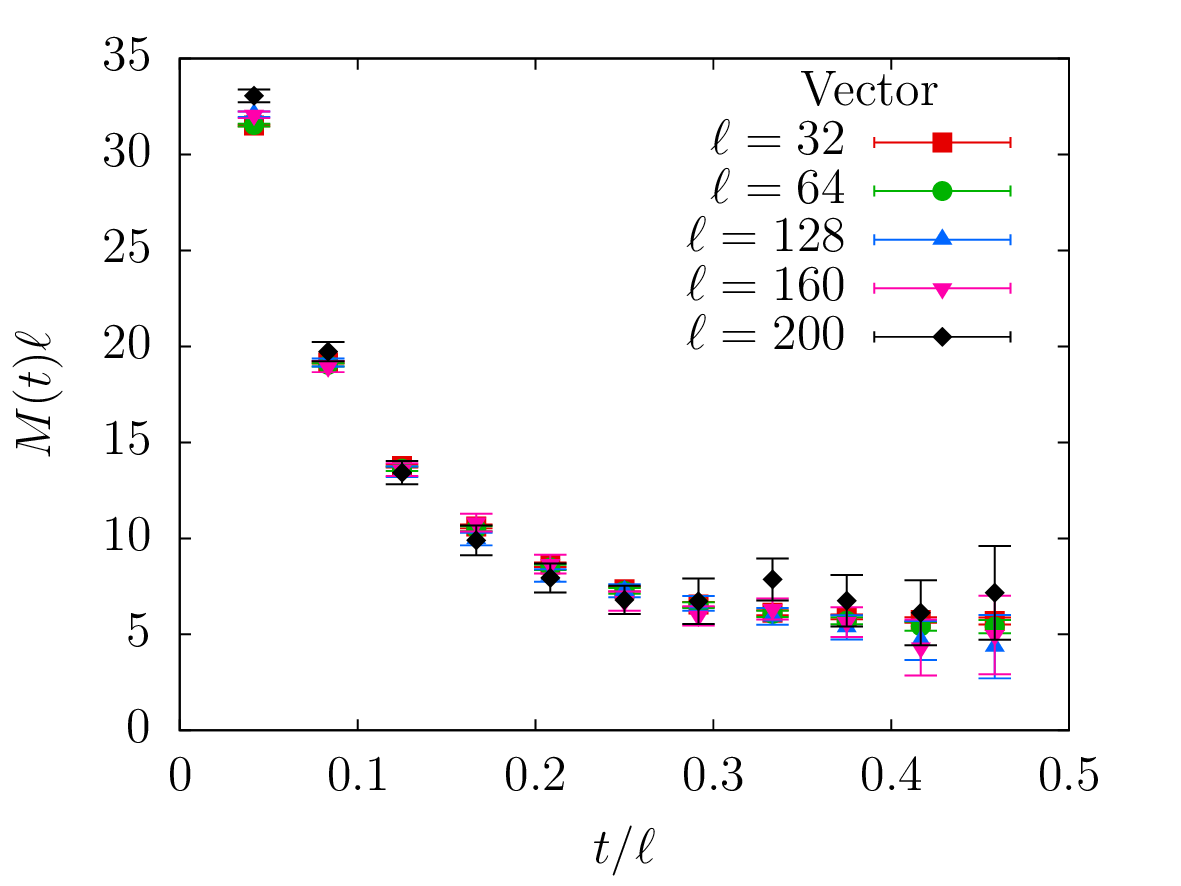}
\caption{The figure shows the dimensionless effective mass $M(t)\ell$
as a function of $\frac{t}{\ell}$ for different $\ell$, represented
by different colored symbols. The top and the bottom panels are for
the scalar and vector respectively, as determined on $L=24$ lattice.
The value of $M\ell$ along the plateau seen in both the panels gives
an upper bound on the mass gap (times $\ell$) present in the scalar
and vector correlators. Since the position of plateau seems to be
independent of $\ell$, $M\sim1/\ell$.
}
\eef{massplot}

In this section, we study the behavior of the scalar and vector
correlators with the aim of lending further support to the scale-invariant 
nature of the $N_f=1$ theory.  As a start, we attempt to
extract a mass using the standard lattice
technique~\cite{gattringer2009quantum} of finding the effective
mass, $M(t)$, from the zero-spatial momentum correlators $G(t)$.
For a correlator in a periodic lattice, one defines the effective
mass $M(t)$ using
\beq
\frac{\cosh\left[M(t)\left(\frac{\ell}{2}-t-\frac{\ell}{L}\right)\right]}
{\cosh\left[M(t)\left(\frac{\ell}{2}-t\right)\right]} = \frac{G\left(t+\frac{\ell}{L}\right)}{G\left(t\right)}.
\eeq{efflat}
If the solution, $M(t)$, to the above equation becomes essentially
independent of $t$ for $0 << t \le \frac{\ell}{2}$, then this
$t$-independent value, $M$, can be used as an upper bound to the
lowest state that contributes to the correlator, $G(t)$. The results
for the scalar and vector effective masses are shown in \fgn{massplot}.
One should note that we have plotted the effective mass times the
box size, $M(t)\ell$, on the $y$-axis.  There is reasonable evidence
for $M(t)\ell$ approaching a limit for large $t$. A striking
observation is that $M \ell$ is essentially independent of the box
size $\ell$. This indicates that the upper bound on the mass $M$
in physical units approaches zero in the infinite volume limit for
both the scalar and vector correlators~\footnote{A fit to both the
scalar and vector correlators with two massive states resulted in
both of the masses approaching zero in the infinite volume.}.
Therefore, there is no mass gap in these two sectors of the theory.
One could take the point of view that the value of $M$ in the plateau
of the effective mass plot is actually a mass gap at finite $\ell$.
In such a case, $M\sim\ell^{-1}$ behavior could be explained as the
standard hyper-scaling relation for a scale-invariant theory. As
for a larger conformal invariance is concerned, such a mass gap
could arise by the explicit breaking of conformal symmetry by the
finite box size, as shown in two-dimensions~\cite{francesco1999conformal}.

\bef
\centering
\includegraphics[scale=0.71]{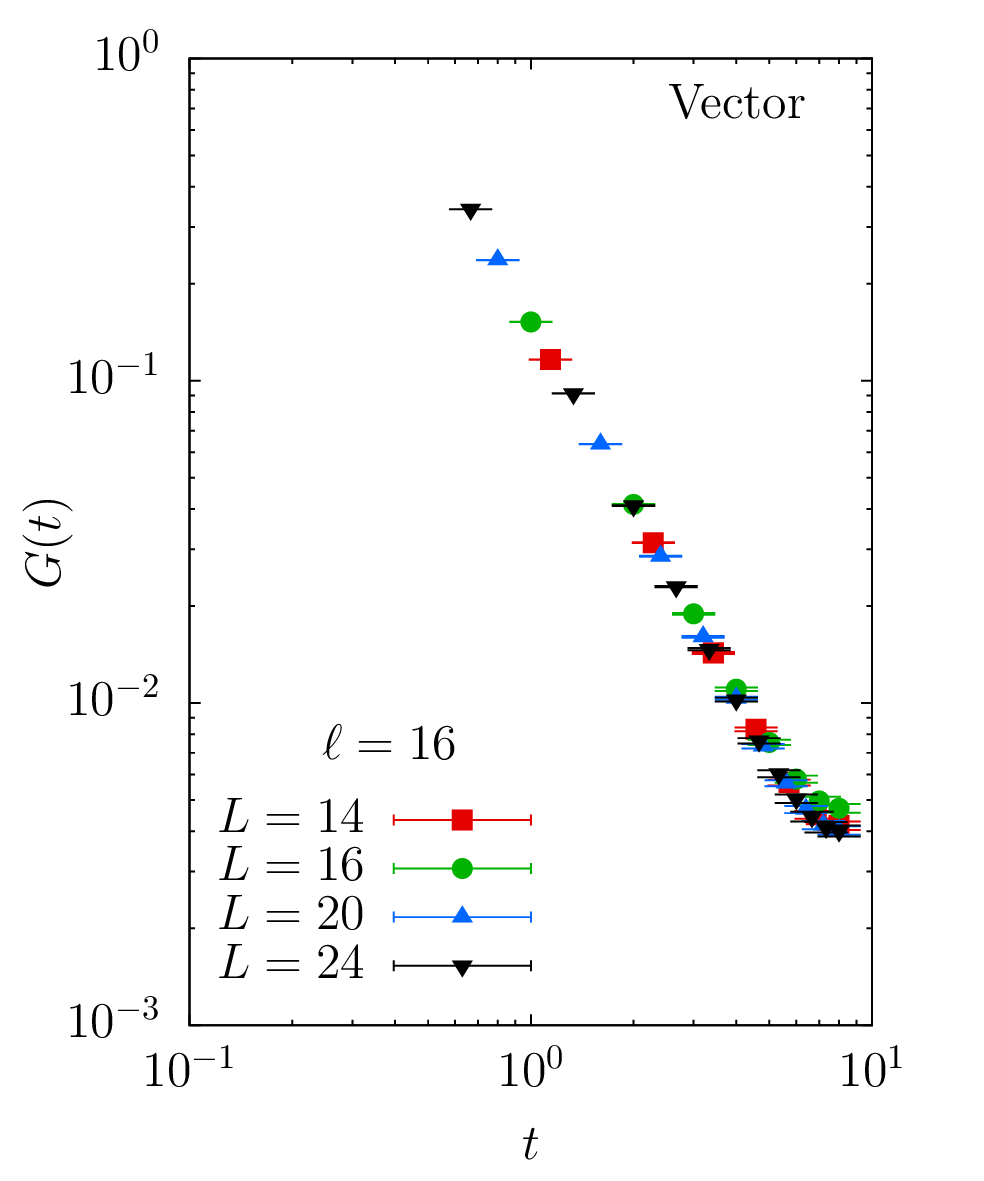}
\includegraphics[scale=0.71]{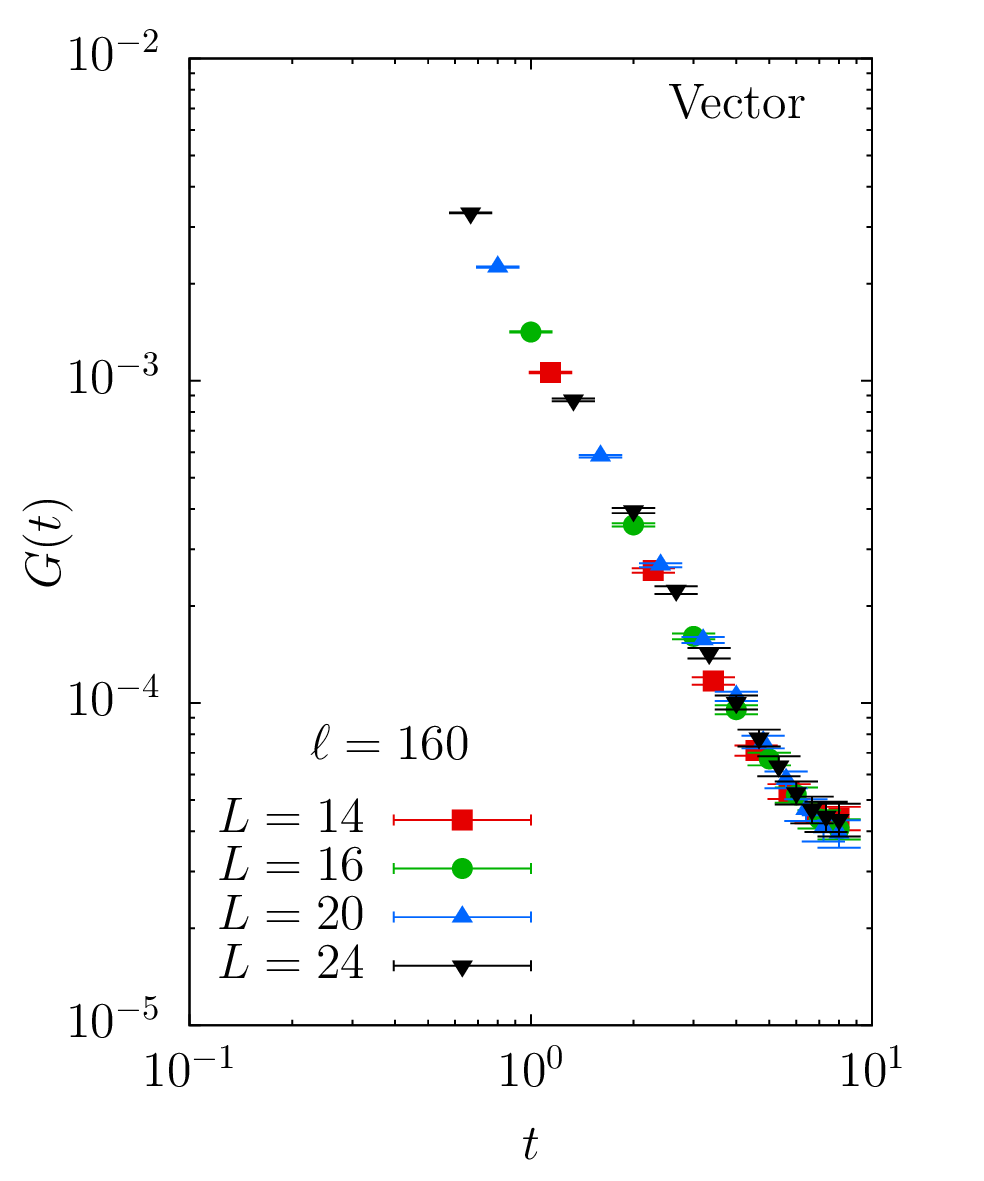}

\includegraphics[scale=0.71]{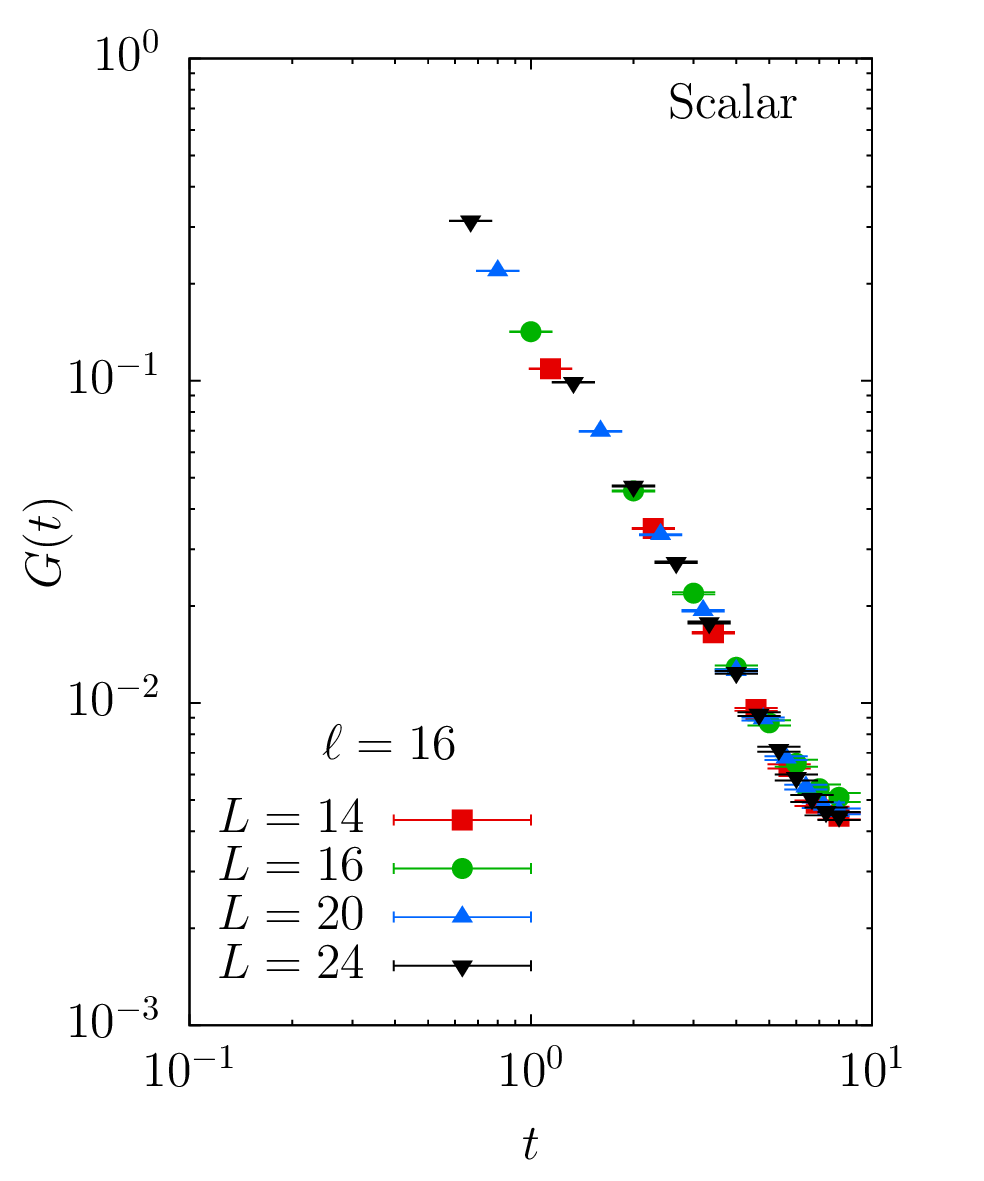}
\includegraphics[scale=0.71]{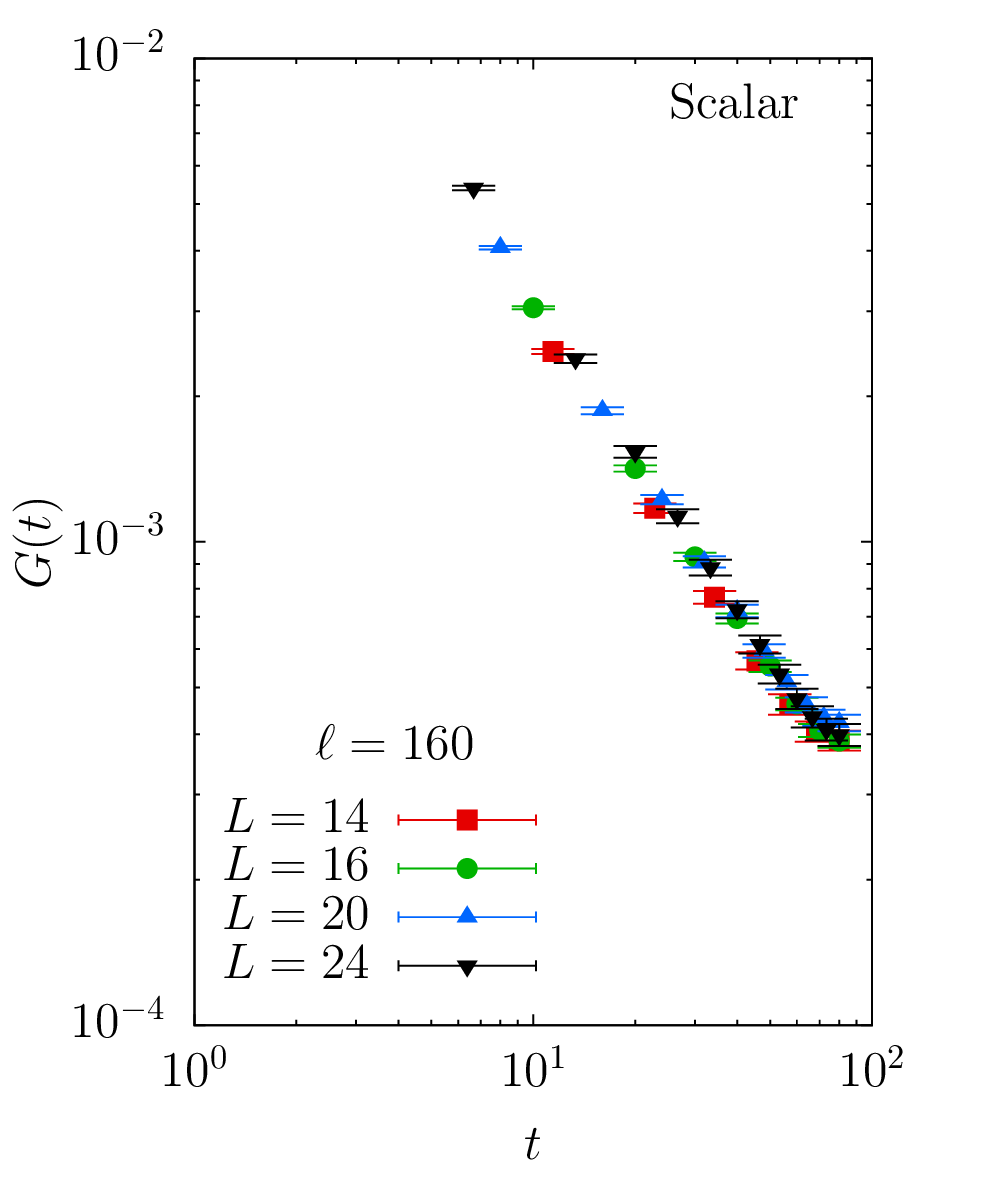}
\caption{Small lattice spacing effects in the correlators. The zero
spatial momentum correlators $G(t)$ are plotted as a function of
physical separation $t$.  The top panels are for the vector and the
bottom ones for the scalar.  The different colored symbols denote
the different $L^3$ lattices used to determine the correlators at fixed
$\ell$.  Lattice effects are small in both finer ($\ell=16$ on the
left panels) and coarser ($\ell=160$ on the right panels) lattices.}
\eef{corvsL}

Next, we look at the correlators themselves to see evidence for a
power-law behavior.  First, we show that the lattice spacing effects
in the scalar and vector correlators are under control in \fgn{corvsL};
the top panels show the vector correlator at different $L$ on two
box sizes representative of small and large $\ell$.  Similar plots
for the scalar are shown in the  bottom panels.  

The brute force way to obtain the correlators at infinite volume
is to take the $L\to\infty$ limit of the correlators at each $\ell$,
and then take $\ell\to\infty$ at each physical $t$. Such a procedure
is not numerically feasible.  Since the lattice artifacts in the
correlators are under control, we can use the correlators determined
at different $\ell$ on the same $L^3$ lattice to scan a wide range
of physical $t$. This is possible provided the $\ell$ dependence
of the correlators at fixed $t$ are small.  Such a reconstruction
of infinite volume correlators for the scalar and vector over a
range of $t$ covering three orders of magnitude are shown in
\fgn{corvst}. We do see a clear approach to the infinite volume
limit at a fixed $t$, after the fact. The vector correlator shows
a clean power law behavior, while the scalar does not. However, the
scalar correlator in the log-log plot in \fgn{corvst} is concave
up, which again rules out the presence of a mass gap because
an exponential on a log-log plot is concave down.  Thus, we are
left with the possibility that the leading scaling behavior for the
scalar would set in at even larger values of physical $t$.

\bef
\centering
\includegraphics[scale=0.6]{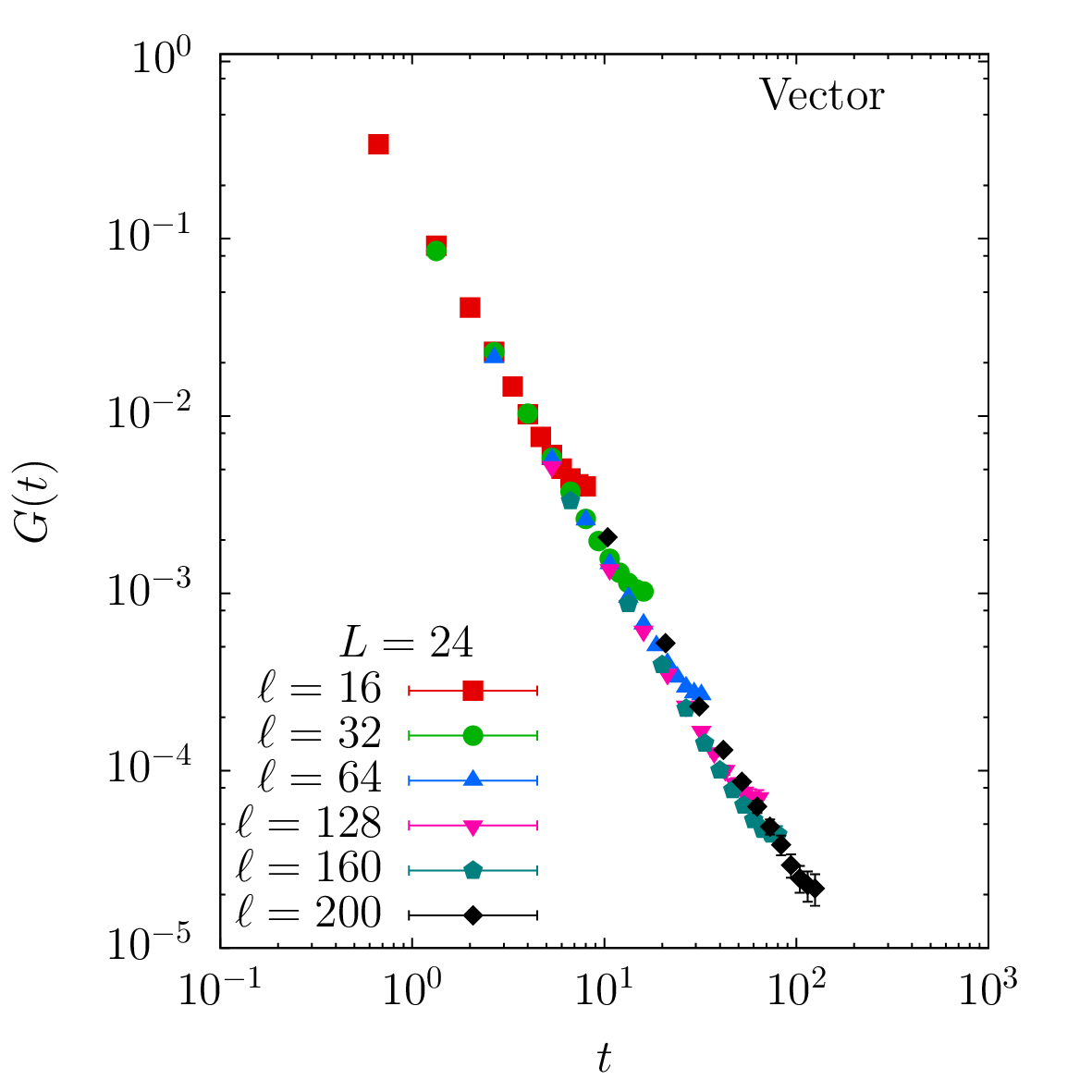}
\includegraphics[scale=0.6]{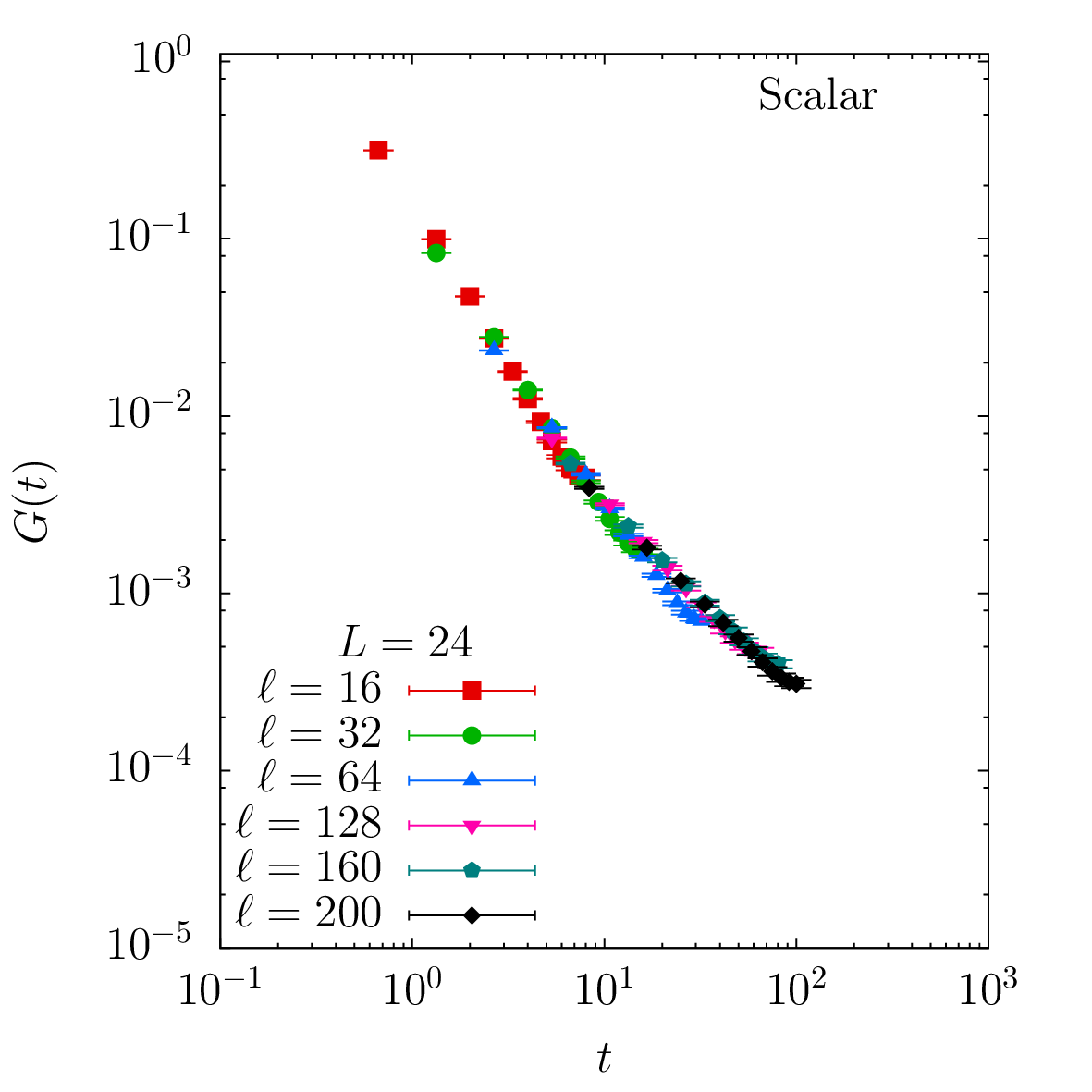}
\caption{The figures show the vector and scalar correlators $G$ as
a function of physical separation $t$, by putting together the data
from various $\ell$, on $L=24$ lattice.  The different colored
symbols correspond to various representative $\ell$. (Left) Vector
shows a power-law behavior as a function of $t$. The small deviations
are due to effect of periodicity at finite $\ell$. (Right) Scalar
does not show a simple power-law, nevertheless massless, as seen
by the concave-up nature of the correlator in the log-log plot.
}
\eef{corvst}

\bef
\centering
\includegraphics[scale=0.6]{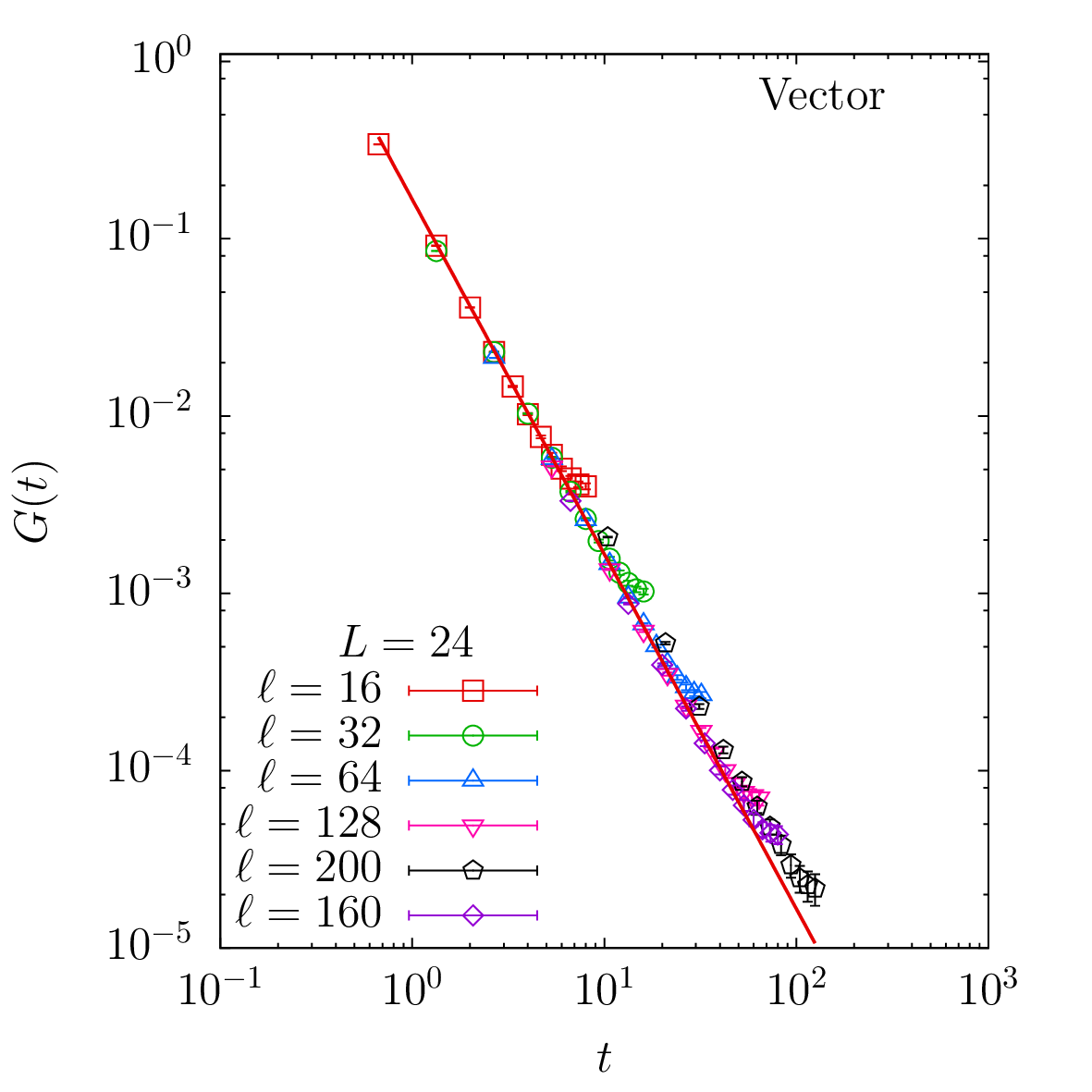}
\includegraphics[scale=0.6]{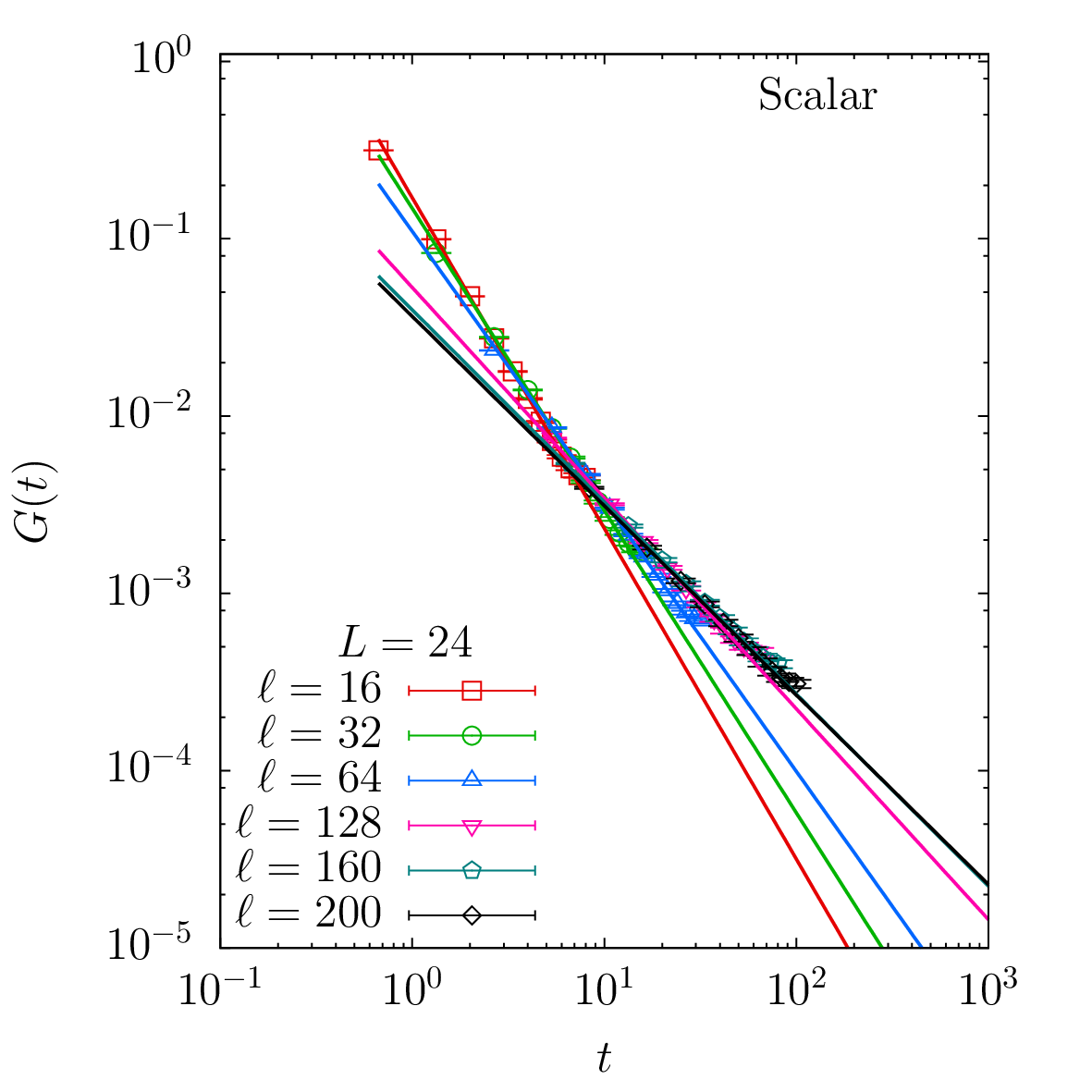}
\caption{
The tangents, which have a slope $d\log(G)/d\log(t)$, are shown
along with the correlator data, same as the ones shown in \fgn{corvst}.
On the left panel, the red solid line corresponds to a $t^{-2}$
power-law. On the right panel, the different colored lines are the
tangents determined at various $t=t_o(\ell)\equiv\ell/L$ using the
finite-$\ell$ scalar correlator data on $L=24$ lattice, which are
represented by symbols of the same color.}
\eef{corslope}

Before we further explore the possible power-law behavior at
larger separations $t$, we digress to consider the expected behavior 
of the zero spatial momentum correlator of a primary operator in a conformal
field theory. Consider the power-law correlation function,
\be 
G_\Delta(x,y,t) \propto\frac{1}{\left(x^2+y^2+t^2\right)^\Delta},
\ee 
corresponding to a primary operator of conformal dimension $\Delta$
in a CFT. Its zero momentum correlation function
will behave as
\be G_\Delta(t) = \int dx dy\  G_\Delta(x,y,t) \propto
\frac{1}{t^k}\qquad\text{where}\qquad k=2(\Delta-1).
\label{powcorr}
\ee
Both the scalar and vector bilinears have an engineering mass
dimension equal to $2$.  Since the vector bilinear is associated
with a conserved current, it will not acquire any anomalous dimension.
Therefore, $\Delta_V=2$, and we should find
\be
G_V(t) \propto \frac{1}{t^2}.
\ee
Since mass acquires an anomalous dimension $1+\gamma_m$, the scalar
bilinear will also acquire an anomalous dimension such that the sum
of the dimensions is equal to $3$. Therefore
\be
\Delta_\Sigma = 2 - \gamma_m.
\label{deltasig}
\ee
We have shown in the previous section that $\gamma_m =1.0 \pm 0.2$,
which along with \eqn{powcorr} and \eqn{deltasig}, suggests that
\be
\Delta_\Sigma = 1.0 \pm 0.2\qquad\text{and}\qquad G_\Sigma(t) \propto \frac{1}{t^{0.0\pm 0.4}}.
\label{kpred}
\ee
Since the correlator has to vanish as $t\to\infty$, we see that
$\gamma_m=1.0$ is marginal for power law behavior which also follows
from unitarity constraints in CFTs~\cite{Mack:1975je}.

\bef
\centering
\includegraphics[scale=0.7]{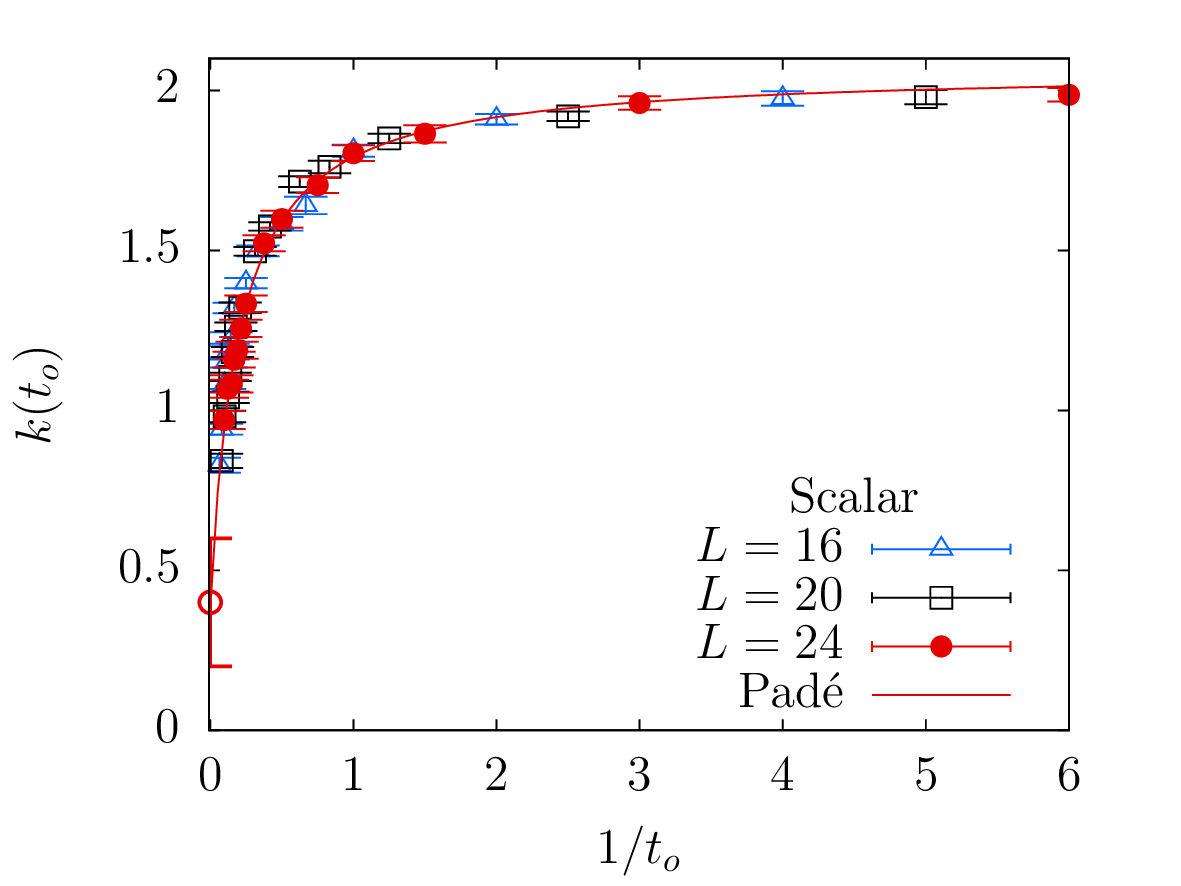}
\caption{ The scale-dependent exponent $k(t)$ for the scalar. The
figure shows the behavior of  $k(t_o)$ defined as
$d\log\left(G(t)\right)/d\log(t)$ determined at various
$t=t_o(\ell)\equiv\ell/L$ using the various finite-$\ell$ scalar
correlator data.  The different colored symbols correspond to
different $L$. At smaller $t_o$, the value of $k$ seems to flow to
2. At larger $t_o$, it keeps decreasing and the value it approaches
as $t_o\to\infty$ is the power-law exponent corresponding to the
infra-red fixed point.  By using a [1/1] Pad\'e in $\tanh(1/t_o)$,
shown by the red solid line, we estimate the infinite $t$ limit of
$k$ to be 0.4(2).  This corresponds to a mass anomalous dimension
$\gamma_m=0.8(1)$.  }
\eef{slopesc}

The vector correlator is shown on the left panel of \fgn{corslope}.
It exhibits a clear power-law behavior over the entire range of the
plot --- $G(t)\sim t^{-2}$ (shown as a red solid line in \fgn{corslope})
describes the data well.  Thus the vector indeed does not get an
anomalous dimension.  This might be non-trivial in light of
Ref~\cite{Collins:2005nj} which was used recently in~\cite{Janssen:2016nrm}
to argue for a possible phase with spontaneously broken Lorentz
symmetry.

The scalar correlator is shown on the the right panel of \fgn{corslope}.
It does not show a simple power law behavior over the entire range.
In particular, the behavior at small $t$ is quite different from
the behavior at large $t$.  In order to estimate the asymptotic
behavior, we use the following strategy.  We numerically estimate
the tangent, $k(t)$, on the log-log plot at various values of $t$.
We have already shown that the $L=24$ data can be assumed to be the
continuum result to a good accuracy.  In any region of $t$ in
\fgn{corslope}, data from multiple $\ell$ overlap. Given a fixed
value of $\ell$, we have a set of $t_n=\frac{\ell(n+1)}{L}$ values
that appear on the plot. We use $n=0,1,2,\ldots,L/4$ and fit a straight line
to the data at a fixed $\ell$ and call that as the tangent, $k(t_0)$.
These are the various colored lines in \fgn{corslope} that correspond
to tangents determined using the correlator data from $\ell$,
represented by the same color.

In \fgn{slopesc}, we show this slope $k$ as a function of $t_o$.
We confirm our earlier statement that the $L=24$ data describes the
continuum quite well by showing that the results from three different
lattice spacings lie on the same curve.  The short-distance behavior
of the correlator is governed by an exponent $k=2$. As $t$ is
increased, the exponent decreases.  One could interpret this as a
flow from the trivial UV fixed point into a non-trivial infra-red
fixed point as the length-scale $t$ is increased. Equivalently, the
bilinear scalar operator is not a simple primary operator in a
conformal field theory but one primary operator dominates the long
distance behavior.  Using a $[1/1]$ Pad\'e approximant in $\tanh(1/t_o)$,
we extrapolate $k(t)$ to its $t\to\infty$ infra-red value.  We
estimate $k(\infty)=0.4(2)$. This corresponds to an estimate of the anomalous
dimension of mass from the scalar correlator, $\gamma_m=0.8(1)$.
This is in good agreement with the one obtained in \scn{massdim}.
Our estimate of the mass anomalous dimension from the correlator
excludes $\gamma_m=1$ since the correlator approaches zero as
$t\to\infty$.

\section{Conclusions}

We have performed a numerical investigation of three dimensional
QED coupled to two flavors of two-component massless fermions while
preserving parity.  We used overlap fermion which preserves the
full U$(2)$ symmetry away from the continuum limit. We extracted
physical quantities on a three-dimensional continuum torus of size
$\ell^3$ by studying the continuum limit at a fixed $\ell$. By
studying the finite-size scaling of the low-lying eigenvalues of
the massless anti-Hermitian overlap Dirac operator, we confirmed
the absence of a bilinear condensate that was previously established
using Wilson-Dirac fermions~\cite{Karthik:2015sgq}.  This enabled
us to obtain a value for the mass anomalous dimension, namely,
$\gamma_m=1.0\pm 0.2$, which is at the upper edge of the allowed
value for a conformal field theory. The eigenvectors associated
with the low lying eigenvalues of the anti-Hermitian overlap Dirac
operator showed critical behavior in the sense of a metal-insulator
transition. The scaling behavior of the inverse participation ratio
(IPR) of the associated eigenvectors and the linear behavior of
number variance of the low lying eigenvalues were consistent with
critical behavior.  Our analysis of the scalar and vector correlators
showed that there is no mass gap in these sectors of the theory. 
The power law behavior of the vector correlator
was consistent with the vector current being conserved. The asymptotic
power law behavior of the scalar correlator resulted in an independent
estimate of the mass anomalous dimension, namely, $\gamma_m=0.8(1)$.

The analysis performed in this paper along with the results
in~\cite{Karthik:2015sgq} suggests that three dimensional QED with
$2N_f$ number of two component massless fermions is scale invariant
for $N_f \ge 1$ when ones uses a regularization that preserves
parity.  In order to consider theories where one has a phase where
scale invariance is broken, we plan to extend our studies to
non-abelian gauge theories in three dimensions.  As a start, we are
currently studying three-dimensional SU$(N)$ gauge theories in the
large $N$ limit where fermions are quenched, provided parity is
preserved. Preliminary numerical studies~\cite{nikhil} suggest that
there is a non-zero bilinear condensate in this limit. The natural
direction we plan to pursue is to map out the phase transition in
the $N$-$N_f$ plane that separates a scale invariant phase from one
where there is a bilinear condensate. Another direction we plan to
pursue is to consider U$(1)$ gauge fields in four dimensions coupled
to fermions in three dimensions. The extent of the fourth direction
changes the gauge action from the limit considered in this paper
where the extent of the fourth direction was set to zero. This is
in the spirit of what one is interested in condensed matter physics
and it is possible that scale invariance is broken if the fourth
direction is large enough.  We also plan to perform numerical studies
in this direction.

\acknowledgments
All computations in this paper were made on the JLAB computing
clusters under a class B project.  The authors acknowledge partial
support by the NSF under grant number PHY-1205396 and PHY-1515446.
We thank Dam Son and Igor Klebanov for useful discussions.

\appendix
\section{Generating functional}\label{sec:genfun}
We develop the basic formula for the generating function
following~\cite{Narayanan:1994gw}. Although the technical details
are not new, the final result shows the explicit form of the
propagator for a two-component fermion.  Let
\be
Z(\eta,\bar\eta) = \langle 0-| \exp\left[ \bar\eta b + a^\dagger\eta \right]|0+\rangle.
\ee
We define new sets of creation operators as
\be
r_+^\dagger = a^\dagger \frac{1+V}{2} \mathcal{R} + b^\dagger \frac{1-V}{2} \mathcal{R};\qquad l_-^\dagger = b^\dagger \mathcal{R},\label{rplm}
\ee
and new sets of annihilation operators as
\be
l_+ = \mathcal{R}^\dagger \frac{1-V^\dagger}{2} a + \mathcal{R}^\dagger\frac{1+V^\dagger}{2} b;\qquad r_- = \mathcal{R}^\dagger a.\label{lprm}
\ee
It follows from \eqn{evhpmr} and \eqn{evhmmr} that
\be
r_+^\dagger |0+\rangle =0;\qquad l_+|0+\rangle =0;\qquad \langle 0-| r_-=0;\qquad \langle 0-|l_-^\dagger =0.
\ee
Inverting \eqn{rplm} and \eqn{lprm}, we arrive at
\be
b^\dagger = l_-^\dagger \mathcal{R}^\dagger; \qquad a^\dagger = r_+^\dagger \mathcal{R}^\dagger \frac{2}{1+V} - l_-^\dagger \mathcal{R}^\dagger A;
\ee
and
\be
a=\mathcal{R}r_-;\qquad b = \frac{2}{1+V^\dagger} \mathcal{R} l_+ + A\mathcal{R} r_-.
\ee
Using the above equations, we can write
\be
\bar\eta b + a^\dagger \eta = Q_+ + Q_-,
\ee
where
\be
Q_+ = r_+^\dagger \mathcal{R}^\dagger \frac{2}{1+V} \eta + \bar\eta \frac{2}{1+V^\dagger} \mathcal{R} l_+;\qquad
Q_- = \bar\eta A\mathcal{R} r_- -l_-^\dagger \mathcal{R}^\dagger A \eta.
\ee
This split is equivalent to
\be
Q_+ = a^\dagger \eta + b^\dagger A\eta - \bar\eta A a +\bar\eta b;\qquad
Q_- = - b^\dagger A\eta + \bar\eta A a.
\ee
From the canonical anti-commutation relations, it follows that
\be
[Q_+,Q_-] = -2\bar\eta A \eta.
\ee
Therefore, we have
\bea
Z(\eta,\bar\eta) &=& \langle 0-| \exp [ Q_+ + Q_-] | 0 + \rangle 
\cr &=& \exp\left ( \frac{1}{2} [ Q_+,Q_-]\right) \langle 0 - | e^{Q_-} e^{Q_+} |0+\rangle\cr
&=& \exp\left ( \frac{1}{2} [ Q_+,Q_-]\right) \langle 0 -  |0+\rangle\cr
&=& \exp [ -\bar \eta A \eta] \det \frac{1+V}{2}.
\eea
This result will be used in \apx{z2mass}.

\section{Introduction of parity invariant mass terms}\label{sec:z2mass}

The partition function for the $N_f=1$ parity-invariant theory with massless fermions is
\be
Z_2=\big\{{}_1\langle 0 - | \otimes {}_2\langle 0 + |\big\}\big\{ | 0+\rangle_1 \otimes | 0-\rangle_2\big\}.
\ee
Parity invariance is ensured by $Z_2\to Z^*_2$ and $1\leftrightarrow 2$.
The generating functional
for $N_f=1$ theory with parity invariant mass terms is
\bea
&&Z_2(\eta_1,\eta_2,\bar\eta_1,\bar\eta_2;m_1,m_2,m_3) =  {}_1\langle 0 - | \otimes {}_2\langle 0 + | \cr
&&\qquad\exp\left[\bar\eta_1 b_1 + a_1^\dagger \eta_1 - \bar\eta_2 a_2 + b_2^\dagger \eta_2
+m_1a_1^\dagger b_1 +m_1 b_2^\dagger a_2 + m_2 a_1^\dagger a_2 + m_3 b_2^\dagger b_1\right]\cr
&&\qquad | 0+\rangle_1 \otimes | 0-\rangle_2 .
\eea
Using standard manipulations of converting the mass terms bilinear in creation-annihilation operators by introducing 
auxiliary Grassmann fields, and then using the result from \apx{genfun}, the final result is
\bea
Z_2(\eta_1,\eta_2,\bar\eta_1,\bar\eta_2;m_1,m_2,m_3)
&=& \det \frac{1+V}{2} \det \frac{1+V^\dagger}{2} \det \left [(1+m_1A)(1-m_1A) + m_2 m_3 A^2\right]\cr
&&
\exp \Biggl[ - \begin{pmatrix} \bar\eta_1 & \bar\eta_2\cr\end{pmatrix}
\frac{
\left [ A \mathbb{I}+ A^2 \begin{pmatrix}
-m_1
& m_2 \cr
-m_3 &
m_1 
\end{pmatrix}
\right]
}{(1+m_1A)(1-m_1A) + m_2 m_3 A^2}
\begin{pmatrix} \eta_1 \cr \eta_2 \cr \end{pmatrix} 
\Biggr].
\eea
We proceed to go to a flavor diagonal basis by diagonalizing the mass matrix using 
\be
W = \begin{pmatrix} m_1 + m & m_2 \cr m_3 & m+m_1\cr \end{pmatrix}.
\ee
where 
\be
 m = \sqrt{ m_1^2 - m_2 m_3 }.
 \ee
 The matrix, $W$, is invertible as long as $m_1^2 \ne  m_2 m_3$ and $m_1\ne -1$.
 Defining flavor diagonal sources as
 \be
 \begin{pmatrix} \bar\eta_+ & \bar\eta_-\cr\end{pmatrix} =  
  \begin{pmatrix} \bar\eta_1 & \bar\eta_2\cr\end{pmatrix} W;\qquad
\begin{pmatrix} \eta_+ \cr \eta_- \cr \end{pmatrix} = W^{-1} \begin{pmatrix} \eta_1 \cr \eta_2 \cr \end{pmatrix} ,
\ee
and noting that
\be
(1+mA) (1-mA) = (1+m_1A)(1-m_1A) + m_2m_3 A^2,
\ee
we arrive at \eqn{z2gen}.

\section{Simulation details}
\label{sec:details}
\bet
\begin{center}
\begin{tabular}{|c||c|c|c|c|c|}
\hline
$\ell$ & \multicolumn{5}{|c|}{$m_t$} \\
\cline{2-6}
 & $L=12$ & $L=14$ & $L=16$ & $L=20$ & $L=24$\\
\hline
4 & 0.032(12) & 0.02233(90) & 0.0182(62) & 0.0091(48) & 0.0067(30)\\
8 & 0.029(10) & 0.0178(87) & 0.0145(72) & 0.0085(41) & 0.0041(31)\\
16 & 0.028(11) & 0.0186(67) & 0.0133(49) & 0.0055(32) & 0.0022(21)\\
24 & 0.0326(84) & 0.0185(60) & 0.0137(46) & 0.0068(23) & 0.0033(21)\\ 
32 & 0.0435(81) & 0.0299(71)& 0.0222(40) & 0.0126(19) & 0.0053(20)\\
48 & 0.072(18) & 0.0542(62)& 0.0396(48) & 0.0244(36) & 0.0145(16)\\
64 & 0.104(10) & 0.0750(85)& 0.0635(45) & 0.0401(37) & 0.0267(21)\\
96 & 0.164(15) & 0.133(12)& 0.1123(75) & 0.0790(52) & 0.0574(38)\\
112 & 0.204(17) & 0.165(15)&0.1337(85) &  0.0996(62) & 0.0747(37)\\
128 & 0.229(22) & 0.193(16)& 0.1633(86) & 0.1181(53) & 0.0917(47)\\
144 & 0.255(12) & 0.221(17) & 0.183(12) &0.1396(91) & 0.1084(52)\\
160 & 0.268(21) & 0.242(19) & 0.207(14) &0.1605(78) & 0.1264(65)\\
200 & 0.323(25) & 0.308(28) & 0.254(14) & 0.206(13)  & 0.1659(66)\\
250 & 0.453(30) & 0.334(23) & 0.297(17) & 0.270(16) & 0.214(12)\\
\hline
\end{tabular}
\end{center}
\caption{The list of tuned Wilson mass $m_t$, which minimizes the
lowest eigenvalue of the Wilson-Dirac operator, for various physical
length of the torus $\ell$ and lattice sizes $L$ used in this study.}
\eet{nf1data}

\bef
\centering
\includegraphics[scale=0.7]{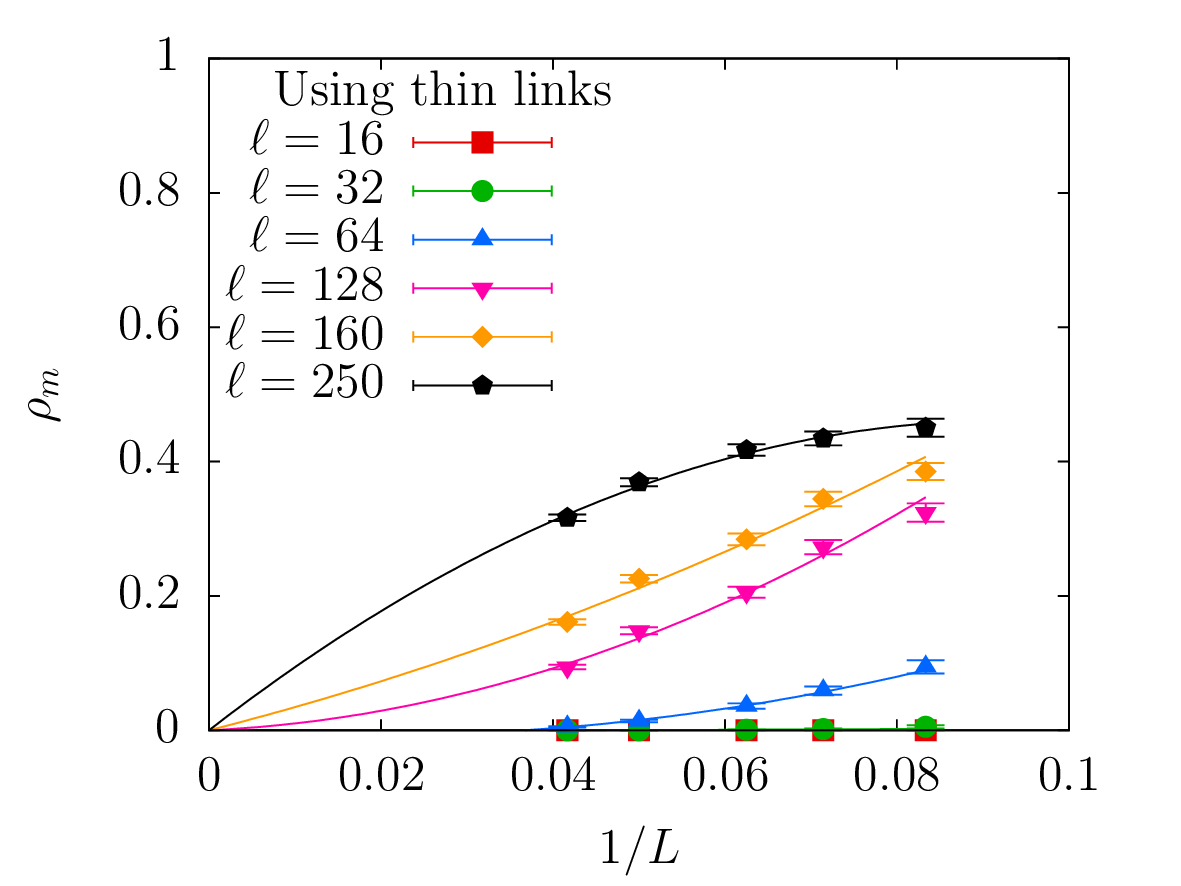}
\includegraphics[scale=0.7]{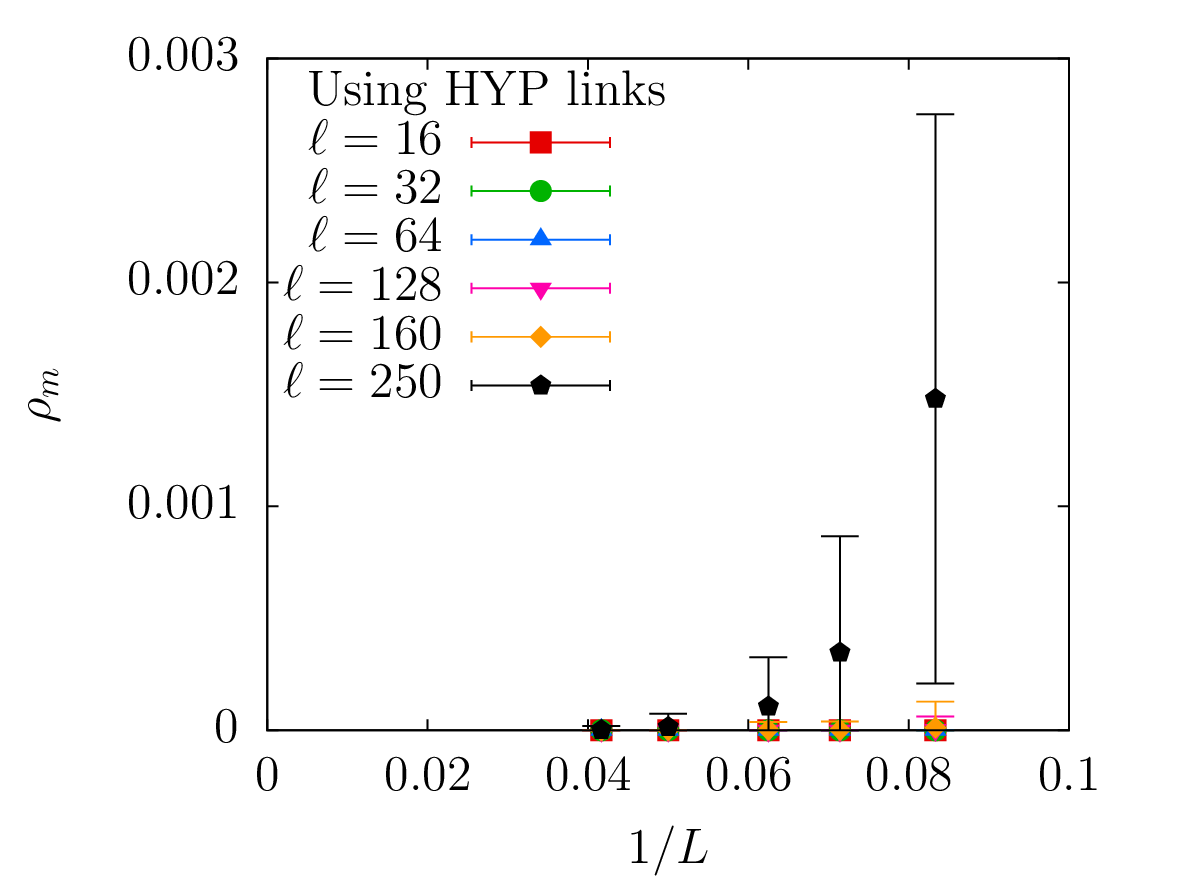}
\caption{The monopole density $\rho_m$ at various $\ell$ as the
continuum limit is approached by taking $L\to\infty$. The left panel
shows the result for the unsmeared, thin gauge-links. The solid
lines show how the density approaches zero using a quadratic $1/L$
extrapolation.  On the the right panel, the result when optimal HYP
smearing is used is shown. With this improvement, the monopole
density is consistent with zero at all simulation points.}
\eef{monopole}

We generated gauge-field configurations using two flavors of dynamical
massless overlap fermions in three-dimensional torus with different
physical extents $\ell$ using $L^3$ lattices. The parameters $\ell$
and $L$ enter the lattice coupling of the non-compact gauge action:
\be
S_g=\frac{L}{\ell}\sum_n\sum_{j< k}^3 \left[\theta_j(n)+\theta_k(n+\hat j)-\theta_j(n+\hat k)-\theta_k(n)\right]^2,
\ee
where $\theta$'s are related to the physical gauge-fields $A_k$ as
$\theta_k=\frac{\ell}{L}A_k$.  We tabulate the set of $\ell$ and
$L$ used in this study in \tbn{nf1data}.  We used the
Sheikhoslami-Wohlert-Wilson-Dirac operator~\cite{Sheikholeslami:1985ij},
adapted to three-dimensions in~\cite{Karthik:2015sgq}, as the kernel
$X$ for the overlap Dirac operator.

As explained in~\cite{Karthik:2015sgq}, we improved the
Sheikhoslami-Wohlert-Wilson-Dirac operator by using one-level HYP
smeared $\theta$'s in the fermion
action~\cite{Hasenfratz:2001hp,Hasenfratz:2007rf}. We used the
optimal smearing parameters $s_1=0.6$ and $s_2=0.5$. Smearing is
essential in our study to explore a range of $\ell$ without the
exorbitant computational cost of using very large $L$. One can see
this by considering the monopole density $\rho_m$ at finite value
of $L$.  In a non-compact U$(1)$ theory, monopoles are not physical
since they are infinite energy objects, hence they are lattice
artifacts. We determined $\rho_m$ for unsmeared as well as optimal
HYP smeared gauge-fields using the procedure outlined
in~\cite{DeGrand:1980eq}.  On the left panel of \fgn{monopole}, we
show that the monopole density in the unsmeared gauge-field indeed
increases with $\ell$ at any finite $L$. At the same time, we also
check that the monopole density vanishes in the continuum limit
$L\to\infty$ at all $\ell$. One could have avoided using smearing,
but lattice artifacts would have been large in the values of $L$
we simulated. On the right panel, we show a similar plot for optimal
HYP smeared gauge fields. Now, we find that even with one-level of
smearing, monopoles are completely removed at all $\ell$. Since the
fermions see only the smeared gauge-fields, it explains why the
fermionic observables exhibit small lattice spacing effects. In
addition to removing monopole-like defects, smearing also results
in a well-defined value of $m_t$ where the smallest eigenvalue of
the Sheikhoslami-Wohlert-Wilson-Dirac operator is minimum. We
tabulate the values of $m_t$, which we use to find the normalization
factor $Z_m$ (refer \eqn{improved}), in \tbn{nf1data}.

We tuned the step-size for the leap-frog evolution at run-time such
that acceptance was at least 80\%. As for the statistics, we generated
$13,000$ to $14,000$ trajectories at each simulation point. Then we
used only configurations separated by an auto-correlation time, as
determined from the smallest eigenvalue $\Lambda_1$. This amounted
to 500 to 1000 independent configurations at all the simulation
points.

\section{HMC force calculation}
\label{sec:force}
In this section, we derive the expression for the fermionic HMC
force for the case of $N_f=1$ massless overlap fermions.  The fermion
force from the pseudo-fermion action in \eqn{pseudo} is
\be
F_\mu(x)=-\frac{\partial S_f}{\partial\theta_\mu(x)}= -\phi^\dagger \frac{\partial\left(C_o^\dagger C_o\right)^{-1}}{\partial\theta_\mu(x)}\phi,
\ee
where the link variable is $U_\mu(x) = e^{i\theta_\mu(x)}$ as defined
in~\cite{Karthik:2015sgq} using the non-compact gauge field. The
fermion force when a smeared gauge-link is used in the fermion
action, as done in this paper, is obtained by the standard chain-rule,
as explicitly worked out in~\cite{Karthik:2015sgq}.

Let us define 
\be
\Psi \equiv \left(C_o^\dagger C_o\right)^{-1} \phi.
\ee
Then
\be
F_\mu(x)= \Psi^\dagger \frac{\partial C_o^\dagger C_o}{\partial \theta_\mu(x)} \Psi.
\label{forcex}
\ee
We use \eqn{ccdag} and $V V^\dagger=1$ to write \eqn{forcex} as
\be
F_\mu(x)=-\frac{1}{2} { \rm Re} \left( \Psi^\dagger V^\prime \Psi\right);\quad V^\prime=\frac{\partial V}{\partial \theta_\mu(x)}.
\ee
$V$ can be computed approximately using
\be
V=X \sum_{k=1}^n \frac{r_k}{X^\dagger X+p_k}.
\ee
We use $n=21$ in our computations. Defining
\be
Y_k \equiv \frac{1}{X^\dagger X + p_k} X^\dagger \Psi; \quad
Z_k \equiv \frac{r_k}{X^\dagger X + p_k} \Psi; \quad
\tilde{Y}_k \equiv X Y_k;\quad
\tilde{Z}_k \equiv X Z_k,
\ee
the fermion force becomes
\be
F_\mu(x) = -\frac{1}{2} { \rm Re} \left\{ \Psi^\dagger X^\prime\left(\sum_{k=1}^n Z_k\right) - \sum_{k=1}^n\left(\tilde{Y}_k^\dagger X^\prime Z_k + \tilde{Z}^\dagger_k X^\prime Y_k\right)\right\},
\ee
where
\be
X^\prime = \frac{\partial X}{\partial \theta_\mu(x)}.
\ee
As explained in \apx{details}, we used Sheikhoslami-Wohlert-Wilson-Dirac
operator in place of the unimproved Wilson Dirac operator, $X$.

\bef
\centering
\includegraphics[scale=0.7]{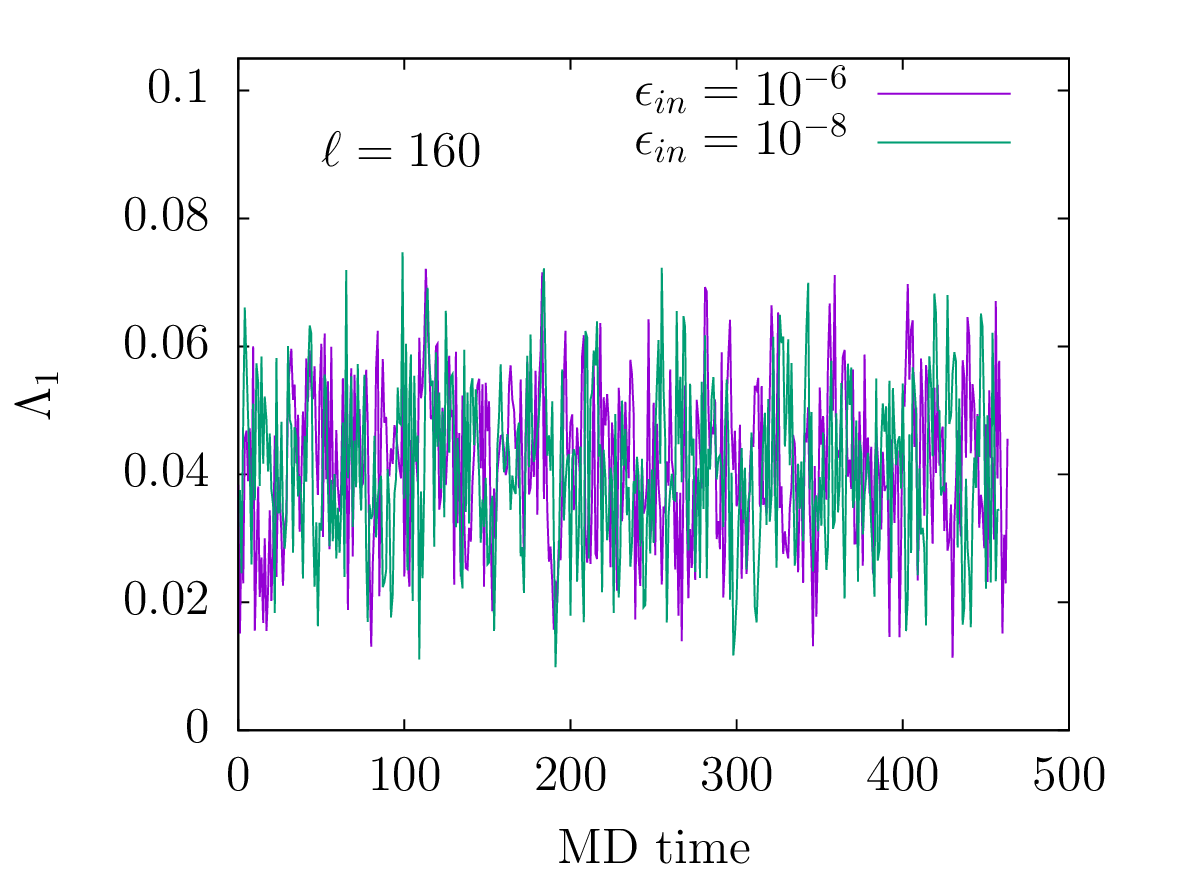}
\caption{Run-time history of the lowest eigenvalue using two different
stopping criterion for the inner CG, $\epsilon_{\rm in}=10^{-6}$
and $10^{-8}$.  The corresponding stopping criterion for the outer
CG were $100\epsilon_{\rm in}$.}
\eef{history}

As is well known, there are two kinds of inversions that enter the
dynamical overlap simulation --- the ones that require the multiple
shifted inversion of $X^\dagger X$ which we refer to as ``inner
CG", and the inversion of $C_o^\dagger C_o$ which we call the ``outer
CG", and it again involves a nested inner CG. We use a stopping
criterion that the ratio of the norm of the residue to the norm of
the solution vector to be less than $\epsilon$. For the inversions
required along the molecular dynamics trajectory, we used a stopping
criterion $\epsilon_{\rm in}=10^{-6}$ for the inner CG, and a
stopping criterion $\epsilon_{\rm out}=10^{-4}$ for the outer CG.
We used a more stringent stopping criterion of $\epsilon_{\rm
in}=10^{-8}$ and $\epsilon_{\rm out}=10^{-6}$ for the inversions
required in the computation of fermion action used in the accept-reject
step. In \fgn{history}, we compare the run-time histories of the
smallest eigenvalue of the overlap operator at $\ell=160$ on $L=14$
lattice. Using a starting thermalized configuration, one of the
runs was made using $(\epsilon_{\rm in}=10^{-6}, \epsilon_{\rm
out}=10^{-4})$  and another with $(\epsilon_{\rm in}=10^{-8},
\epsilon_{\rm out}=10^{-6})$. We find that it is sufficient to use
the less stringent stopping criterion.

\bibliography{biblio}
\end{document}